\documentclass[review]{elsarticle}

\usepackage{lineno,hyperref}
\usepackage{amsmath}
\usepackage{amssymb}
\usepackage[]{algorithm2e}
\usepackage{amssymb}
\usepackage{graphicx}
\usepackage{subcaption}
\usepackage{xcolor}
\usepackage{placeins}
\journal{Journal of Elsevier Signal Processing}

\begin{document}
	
	\begin{frontmatter}
		
		\title{A Nonlinear Acceleration Method for Iterative Algorithms}
		\author[mymainaddress]{Mahdi Shamsi}
		
		\author{Mahmoud Ghandi}
		
		\author[mymainaddress]{Farokh Marvasti}
		
		
		
		\address[mymainaddress]{Multimedia and Signal processing Lab (MSL),
			Advanced Communications Research Institute (ACRI),
			EE Department of Sharif University of Technology, Tehran, I.R.Iran}
		
		\begin{abstract}
			Iterative methods have led to better understanding and solving {\color{black}problems such as missing sampling, deconvolution, inverse systems, impulsive and Salt and Pepper noise removal problems.} However, the challenges such as the speed of convergence and or the accuracy of the answer still remain. In order to improve the existing iterative algorithms, a non-linear method is discussed in this paper. The mentioned method is analyzed from different aspects, including its convergence and its ability to accelerate recursive algorithms.{\color{black} We 
				show that} this method is capable of improving Iterative Method (IM) as a non-uniform sampling reconstruction algorithm and some iterative sparse recovery algorithms such as {\color{black}  Iterative Reweighted Least Squares (IRLS), Iterative Method with Adaptive Thresholding (IMAT), Smoothed $\ell_0$ (SL0) and Alternating Direction Method of Multipliers (ADMM) for solving LASSO problems family (including Lasso itself, Lasso-LSQR and group-Lasso)}. It is also capable of both accelerating and stabilizing the well-known Chebyshev Acceleration (CA) method. Furthermore, the proposed algorithm can extend the stability range by reducing the sensitivity of iterative algorithms to the changes of adaptation rate.
		\end{abstract}
		\begin{keyword}
			Non-Linear Acceleration; Iterative Methods; Sparse Recovery; Acceleration Methods; IMAT; LASSO.
		\end{keyword}
		
	\end{frontmatter}
	
	
	\section{Introduction}
	\label{sec:intro}
	Ideally, the solution of a problem is obtained either in a closed form or by an analytical approach. However, this perspective is not applicable to most cases since a closed form does not necessarily exist and even if it does, it might be too complicated to achieve. One way to approach such problems is using iterative algorithms. Despite being applicable to many cases, these algorithms have their own disadvantages. For instance, they become more complex as the number of iterations increases. Besides, their convergence/stability should be considered as well.
	
	In order to accelerate iterative algorithms, many different methods have been proposed. Polynomial acceleration techniques are used to iteratively solve a large set of linear equations \cite{grochenig1993acceleration,saad1984chebyshev}. The Chebyshev Algorithm (CA) for example, is a polynomial acceleration method that has been introduced to speed up the convergence rate of frame algorithms. Conjugate Gradient (CG) methods are amongst the most useful algorithms for solving optimization problems and can be simply adapted to accelerate nonlinear iterative methods such as CG-Iterative Hard Thresholding (IHT) \cite{grochenig1993acceleration,nocedal2006conjugate,blanchard2015cgiht}.
	Accelerating methods are mostly proposed in order to increase the convergence rate of iterative algorithms based on their target signals. As a result, each accelerating method is only capable of accelerating a limited number of iterative algorithms.
	
	In this paper, a Non-Linear (NL) acceleration method is used to increase the convergence rate  {\color{black}of any iterative} algorithms. The  {\color{black}proposed} method is also capable of stabilizing some diverging algorithms. Previously, a similar idea but with a different point of view was used to accelerate an Iterative Method (IM) \cite{feichtinger1994theory,marvasti1989iterative} for non-uniform missing samples recovery problem regarding 1-D Low-Pass (LP) signals\cite{ghandi2008some}. Before that, Aitken used this method to accelerate the rate of converging sequences only \cite{aitken1927xxv}.
	
	The NL method is capable of increasing the convergence rate of optimization algorithms. Iterative methods are widely used in gradient-based optimization algorithms such as AdaGrad for high dimensional sparse gradients \cite{duchi2011adaptive}, RMSprop for non-stationary and real-time scenarios \cite{tieleman2012lecture} and AdaMax for a combined case of online and sparse gradient-based optimization problems \cite{kingma2014adam}.
	
	{\color{black}Least Absolute Shrinkage and Selection Operator (LASSO) have been popularized as a regularized Least-Squares Estimation (LSE) problem \cite{tibshirani1996regression, hastieelements} which induces some sparsity to the LS solution. Group-Lasso was introduced to  allow predefined groups of covariates to be selected into or out of a model together \cite{yuan2006model}. LSQR was proposed to solve sparse linear equations and sparse least squares; it can be used to improve LASSO solving algorithms in the case of ill-conditioned measurement matrices \cite{paige1982lsqr}. There are lots of algorithms for solving the Lasso problem such as Alternating Direction Method of Multipliers (ADMM) which can iteratively solve LASSO problems family\cite{wahlberg2012admm,boyd2011distributed}.}
	
	In this paper, after stabilizing the NL method, we extend it to accelerating image recovery algorithms as well as sparse recovery methods. We then study its interesting capability of stabilizing diverging algorithms. 
	
	In a nutshell, the present study consists of (1) improving previous works in order to accelerate and stabilize the IM with higher values of relaxation parameter, (2) accelerating iterative image recovery algorithms and (3) applying the NL method in order to accelerate iterative sparse recovery algorithms. The convergence of the proposed method is analyzed in three different categories of sub-linear, linear and super-linear converging sequences based on the sign changes of the errors in three successive estimations. These statements are  {\color{black}confirmed} by simulations of various iterative algorithms.
	
	This paper is organized as follows: In Section \ref{sec:Preliminaries} we briefly review some signal recovery algorithms as well as {\color{black}the CA}. The {\color{black}NL} algorithm is studied in Section \ref{sec:solAlg}. In Section \ref{sec:sim} the simulation results are reported. Finally, in Section \ref{sec:conclusion}, we {\color{black}will} conclude the paper.
	\section{Preliminaries}
	\label{sec:Preliminaries}
	In this section, iterative algorithms are considered as a broad group of problem solving approaches and some of them are reviewed.
	
	The IM was first proposed to compensate for the distortion caused by non-ideal interpolation. By defining G as a distortion operator it is desired to find G-1 to compensate for its distortion. The error operator could be defined as
	\begin{equation}
	E\triangleq I-G\notag
	\end{equation}
	where $I$ is the identity operator. Hence we can write
	\begin{align}
	G^{-1}&=\frac{I}{G}=\frac{I}{I-E}=I+\sum_{n=1}^{\infty}E^n; \parallel E\parallel<1\notag\\
	&\Rightarrow G^{-1}_{_{<K>}}=I+\sum_{n=1}^{K}E^n\notag
	\end{align}
	where $G^{-1}_{_{<K>}}$ is the $k^{th}$ order estimation of $G^{-1}$. It is clear that
	\begin{equation}
	G^{-1}_{_{<K+1>}}=I+E(G^{-1}_{_{<K>}}).\notag
	\end{equation}
	The convergence rate of the IM can be controlled by defining a relaxation parameter such as $\lambda$ in
	\begin{align}
	G^{-1}&=\frac{\lambda I}{\lambda G}=\frac{\lambda I}{I-E_\lambda}; E_\lambda\triangleq I-\lambda G\:,\:\parallel E_\lambda\parallel<1\notag\\
	&\Rightarrow G^{-1}_{_{<K,\lambda>}}=I+\sum_{n=1}^{K}{E_\lambda}^n\notag
	\end{align}
	which can be recursively implemented by the equation below:
	\begin{equation}
	x_{k} =\lambda (x_{0}-G(x_{k-1}))+x_{k-1}
	\label{eq:IM}
	\end{equation}
	where $x_{k}$ is the $k^{th}$ estimated signal.
	It has been proved that the IM leads to the pseudo-inverse solution and that the convergence (in the sense of stability and speed) can be controlled by tuning the relaxation parameter ($\lambda$) \cite{amini2007reconstruction}. The IM is suitable for reconstructing band-limited signals and by choosing a proper $G$,it can be used as a non-uniform missing sample recovery algorithm \cite{marvasti2012nonuniform}.
	
	Most signals are not band-limited. However, they can be sparse in some other domains. Sparse recovery is a broad problem in the literature of signal recovery. Assuming that a given signal is sparse in a specific domain, it can be perfectly reconstructed even with a limited number of observations.  The main problem in sparse recovery is the minimization of an $\ell_0$ semi-norm minimization ($P_0$ problem) \cite{donoho2006stable}. Because of non-convexity, the $P_0$ problem is usually substituted with an $\ell_1$ norm minimization problem ($P_1$ problem). , an $\ell_1$ norm minimization ($P_1$ problem) is usually substituted for the $P_0$. It has been shown that under some conditions regarding the signal sparsity number and the observation matrix, the solution of P0 can be obtained by solving $P_1$ \cite{donoho2006stable,donoho2006most}. We have
	\begin{align}
	P_0: \quad \min_s{\| s\|}_0\: ;\: As=b\:,\:
	P_1: \quad \min_s{\| s\|}_1\: ;\: As=b\notag
	\end{align}
	where $A$ and $b$ are the fat observation matrix and the observed signal, respectively.
	
	{\color{black}The method of Iteratively Reweighted Least Squares (IRLS) is used to iteratively solve approximated $P_0$ with a weighted LS problem \cite{chartrand2008iteratively,daubechies2010iteratively}.} Another approach to sparse recovery is approximating the $\ell_0$ semi-norm by a smooth function. Smoothed $\ell_0$ (SL0) method is an iterative sparse recovery method which can be used to approximate the $\ell_0$ semi-norm with a smooth function \cite{mohimani2009fast} such as
	\begin{align}
	f_{\sigma}(s)=  N-\sum\limits_{n=1}^{N} F_{\sigma}(s[n])\:;\:{F_{\sigma}(s[n])}=\exp(-\frac{{|s[n]|}^2}{2\sigma^2})\notag
	\end{align}
	where $s$ is a sparse signal with the length $N$ and $s[n]$ is its $n^{th}$ component. It can be seen that $f_0(s)=\|s\|_0$; by this approximation, the problem can be reduced to an ordinary optimization problem. As a result, its minimum can simply be found using simple minimization methods such as Steepest Descent (SD) method. It should be noted that by assigning a very small value to $\sigma$ the algorithm is trapped in a local minimum \cite{eftekhari2009robust}. In order to escape the local minimum, the algorithm is run for a moderately larger $\sigma$ and after some iterations, the estimated signal is used for initializing the next run of the algorithm with a smaller value of $\sigma$ (it can be reduced by a decreasing factor such as Sigma Decreasing Factor ($SDF$)). This process lasts until the algorithm converges. In order to satisfy the observation constraints, after each gradient step, a projection step is required, as shown in Alg.\ref{alg:sl0}.
	\begin{algorithm}[htb]
		\KwData{$A,b,\sigma_0,M,K,r,\mu_0$}
		\KwResult{$s$ }
		$s=A^+b$\;
		\For {m=0:M-1}{
			$\sigma\leftarrow{r}^{m}\times\sigma_0$\;
			\For{k=0:K-1}
			{$G:$ $s\leftarrow s-\mu_0\nabla f_{\sigma}(s)$\\
				$P:$ $s\leftarrow s-A^+(As-b)$}
		}
		\caption{The SL0 algorithm.}
		\label{alg:sl0}
	\end{algorithm}
	
	Another approach to solve sparse recovery problems is modifying inverse algorithms. In order to use the IM for sparse recovery it first needs to be properly modified. Iterative algorithms such as the IHT –which guarantees suboptimal signal recovery with robustness against additive noise \cite{blumensath2008iterative,blumensath2009iterative}- and Iterative Method with Adaptive Thresholding (IMAT) \cite{marvasti1987unified}, use thresholding (in a specified transform domain) as an approach to sparse recovery. IMAT can lead to faster and better reconstruction performance compared to the IHT. Besides, the IHT needs prior knowledge regarding the sparsity number while it is not necessary for IMAT \cite{azghani2013iterative,azghani2015microwave}.
	IMAT can be interpreted as a modification for the IM. This can be realized by using a threshold function after each iteration of the IM, as opposed to low pass filtering (\ref{eq:IMAT})
	\begin{align}
	x_{k} &=T[\lambda (x_{0}-G(x_{k-1}))+x_{k-1}]
	\label{eq:IMAT}\\
	T(x)&= \begin{cases}
	X=Tr(x)	\\
	Y=\begin{cases}
	X\quad ;\quad  T_k\le|X|\\
	0 \quad ;\quad \text{else} 
	\end{cases}\\
	\text{return} \quad ITr(Y)
	\end{cases}\notag
	\end{align}
	where $T(.)$ is the thresholding function, $Tr(.)$ and $ITr(.)$ are respectively a transformation operator and its inverse. $Tr(.)$ needs to be properly chosen in order to transform the signal to its sparsity domain. One common approach is to reduce the exponential threshold function in each iteration using the equation: $T_k=T_0\mathrm{e}^{-\alpha k}$
	where $T_0$ and $\alpha$ are hyper-parameters used to control the threshold values in each iteration.
	
\begin{figure}[htb!]
		\centering
		\includegraphics[width=\textwidth]{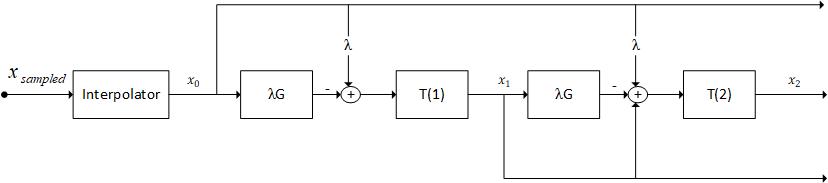}
		\caption{Block Diagram of IMATI algorithm with distortion operator $G$ and relaxation parameter of $\lambda$.}
		\label{fig:IMATI_blck}
	\end{figure}

	In non-uniform sampling, $G$ is defined as the sampling operator. Appropriate modification of $G$ can lead to improving the performance of IMAT. A modified version of IMAT is IMAT with Interpolation (IMATI) in which IMAT is improved by using a sampling operator followed by an interpolator or a smoothing function as the distortion operator, as shown in Fig.\ref{fig:IMATI_blck}.
	
	IMAT and IMATI algorithms can also be simply used for image reconstruction and also 3-D signal recovery \cite{azghani2013iterative}. 
	
	In order to increase the speed of frame algorithms, the CA method can be used. It is represented by the following equations
	\begin{align}
	&x_1=\frac{2}{A+B}x_0,\,\lambda_k=(1-\frac{\rho^2}{4}\lambda_{k-1})^{-1}; \,\text{for}\, k>1:\notag\\
	&x_k=(x_1+x_{k-1}-\frac{2}{A+B}G(x_{k-1})-x_{k-2})\lambda_k+x_{k-2}\notag
	\end{align}
	where $A,B>0$ are the frame bounds which can control the convergence of the algorithm.  Hence, inappropriate selection of these parameters can result in divergence.
	\section{Theory and Calculation}
	\label{sec:solAlg}
	In this section, the importance of the convergence rate and the stability of iterative algorithms are discussed. Even though these two subjects are generally inconsistent, the NL method and its modification are introduced in order to both speed up and stabilize the iterative algorithm.
	
	Assuming $\hat{x}_k[n]$ is the $k^{th}$ estimation of the desired signal at time index $n$, the corresponding recovery error is given by $e_k[n]=\hat{x}_k[n]-x[n]$. Generally, $e_k[n]$ can be written as a proportion of $e_{k-1}[n]$,
	\begin{equation}
	e_k[n]=\alpha_{k-1}[n]\times e_{k-1}[n]\:;\: e_{k-1}[n]\neq 0\notag
	\end{equation}
	where $\alpha_k[n]$ is the coefficient of proportionality. To be concise and in order not to lose generality, we consider three typical estimated signals, such as  $\hat{x}_1$, $\hat{x}_2$ and $\hat{x}_3$, for a specific time index. By assigning the same value to the first two successive $\alpha_k$'s (i.e.,  $\alpha_1=\alpha_2=\alpha$), we can write
	\begin{equation}
	\hat{x}_3-x=\alpha(\hat{x}_2-x)=\alpha^2(\hat{x}_1-x).
	\label{eq:err}
	\end{equation}
	By computing $x$ from (\ref{eq:err}) the following Non-Linear (NL) formula is obtained:
	\begin{equation}
	x_{_{NL}}=\frac{\hat{x}_3\times\hat{x}_1-\hat{x}_2^2}{\hat{x}_3+\hat{x}_1-2\hat{x}_2}.\notag
	\end{equation}
	
	Considering  $\hat{x}_i=x+e_i$ for $i=1,2,3$, it can be deduced from the NL formula that
	\begin{equation}
	x_{_{NL}}=x+e_{_{NL}}\;;\;e_{_{NL}}\triangleq\frac{{e}_3\times{e}_1-{e}_2^2}{{e}_3+{e}_1-2{e}_2}.\notag
	\end{equation}
	
	Due to the sign alternations of $e_i$'s, it can be shown that in some cases $|e_{_{NL}}|$ can be larger than $|e_3|$ -which is probably smaller than both $|e_2|$ and $|e_1|$-. In other words, the error obtained while using the NL algorithm is larger than the errors of the existing estimations. In order to analyze the convergence of the NL method, it should be noted that there exist $2^3=8$ possible cases for the sign alternations of $e_i$'s. Assuming that $sign(e_2)=1$, this number can be reduced to $2^2=4$. Furthermore, there are two cases based on the relative changes of $|\alpha_1|$ and $|\alpha_2|$. The first case is linear convergence (for $|\alpha_1|=|\alpha_2|$) while we consider the second case to serve as the two cases of sub-linear and super-linear convergence (for $|\alpha_1|\neq|\alpha_2|$).
	
	In order to make sure that the estimation does not diverge when divided by zero, the NL method is applied to $\hat{x}_0$, $\hat{x}_1$ and $\hat{x}_2$ \cite{ghandi2008some}. Therefore, by defining
	\begin{align}
	\sigma_1\triangleq\hat{x}_3+\hat{x}_1-2\hat{x}_2\,,\,\sigma_0\triangleq\hat{x}_2+\hat{x}_0-2\hat{x}_1,\notag
	\end{align}
	the Modified NL (MNL) method is obtained and can be represented as follows
	\begin{equation}
	x_{_{MNL}}=\begin{cases}
	\frac{\hat{x}_3\times\hat{x}_1-\hat{x}_2^2}{\hat{x}_3+\hat{x}_1-2\hat{x}_2}\:;\: \vert\sigma_0\vert\leq\vert\sigma_1\vert\\
	
	\frac{\hat{x}_2\times\hat{x}_0-\hat{x}_1^2}{\hat{x}_2+\hat{x}_0-2\hat{x}_1}\:;\: \vert\sigma_0\vert>\vert\sigma_1\vert
	\end{cases}.
	\end{equation}
	\newline
	Apart from that, the error is unavoidable for very small values of  $\sigma_i$ due to the finite precision hardware implementation. Fortunately, the latter divergence occurs only in a very small number of points of the NL estimation and is usually noticeable as high spikes.
	
	Different simple techniques can be used in order to compensate for these issues. One approach could be using some of the linear combinations (weighted averages) of the existing estimations in substitution instead. This can lead to reducing the sign alternations of the errors. Another approach is applying the NL formula to transformed versions of the estimations in another specified domain. Assuming that the signals are bounded, simple methods such as Clipping, Substitution and Smoothing can be used in order to compensate for the undesirable spikes caused by the NL method. For each up-crossing of the foreknown levels (the maximum and the minimum) we either clip the estimated signal or substitute the best existing signal for the reconstructed signal. If the signal is too noisy, the Median Filter (MedFilt) can be used in order to smooth the NL signal. In this article, we use both MedFilt and Clipping.
	
	One interesting feature of the NL formula is its symmetry. For a diverging algorithm we have
	\begin{equation}
	|e_1|<|e_2|<|e_3|.\notag
	\end{equation}
	For any selection of $\hat{x}_1$, $\hat{x}_2$ and $\hat{x}_3$, whether the error is decreasing or increasing, the NL formula leads to the same results since $\hat{x}_3$, $\hat{x}_2$ and $\hat{x}_1$ can be considered to be three successive converging estimations. For a constant rate of changes for three successive errors (either converging or diverging), using the NL method can lead to a perfect signal reconstruction. This property can be used in Watermarking and Steganography.
	\section{Simulation Results and Discussion}
	\label{sec:sim}
	In this section we use the MNL method to accelerate some iterative algorithms. There are so many algorithms which solve the problems recursively. Even though the introduced NL method is capable of improving most of the existing iterative algorithms under some simple assumptions, we focus on recovering the signal from its with missing samples (special case of nonuniform samples). Missing samples occur at random indices with iid Bernoulli distribution with the parameter $p=\text{Loss-Rate (}LR$).
	
	{\color{black}
		In order to illustrate the performance of image recovery algorithms, some Image Quality Assessment (IQA) methods are used. Peak Signal to Noise Ratio (PSNR), Inter-Patch and Intra-Patch based IQA (IIQA) \cite{zhou2015image}, Structural Similarity (SSIM) \cite{wang2004image} (Convex
		SIMilarity (CSIM) as its convex version \cite{javaheri2018robust}), Multi-Scale SSIM (MS-SSIM) \cite{wang2003multi}, Edge Strength Similarity (ESSIM) \cite{zhang2013edge} and Feature Similarity (FSIM) \cite{zhang2011fsim} are used as Full-Reference (FR) IQA methods while Blind/Referenceless Image Spatial Quality Evaluator (BRISQUE) \cite{mittal2011blind,mittal2012no} and Naturalness Image Quality Evaluator (NIQE) \cite{mittal2013making} are used as No-Reference (NR) IQA methods.}
	
	The MNL formula was applied to the IM in order to reconstruct the 1D sampled signals (with the length $L=500$) with a specified Over Sampling Ratio (OSR). The LP signals were generated by filtering the white normal noise using the DFT Filter. In order to achieve fair results, the result of each experiment was averaged over 100 runs. 
	
\begin{figure}[tbh]
		\centering
		\begin{subfigure}[h]{.48\linewidth}
			\includegraphics[width=\textwidth,height=.6\textwidth]{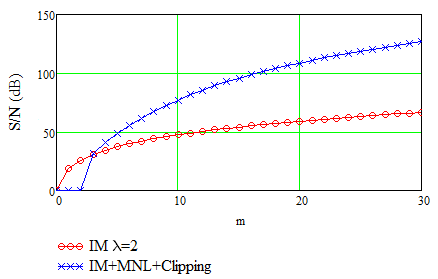}
			\caption{The MNL+Clipping.}
			\label{fig:MNL_IM_1D_spikes}
		\end{subfigure}
		\begin{subfigure}[h]{.48\linewidth}
			\includegraphics[width=\textwidth,height=.6\textwidth]{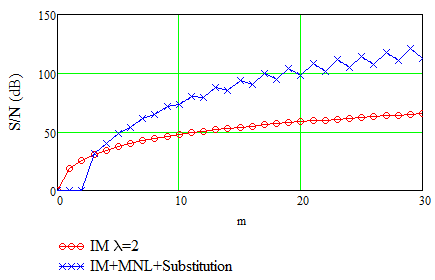}
			\caption{The MNL+Substitution.}
			\label{fig:MNL_IM_1D}
		\end{subfigure}
		\caption{SNR curves of the IM and the MNL, $OSR=8$, $LR=\frac{1}{3}$, $\lambda=2.2$.}
		\label{fig:1D_IM}
	\end{figure}

\begin{figure}[tbh]
		\centering
		\begin{subfigure}[h]{.48\linewidth}
			\includegraphics[width=\textwidth,height=.6\textwidth]{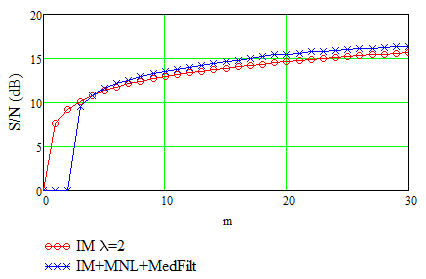}
			\caption{The MNL+MedFilt.}
		\end{subfigure}
		\begin{subfigure}[h]{.48\linewidth}
			\includegraphics[width=\textwidth,height=.6\textwidth]{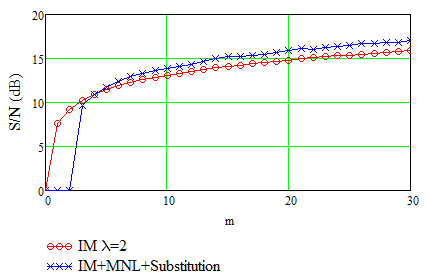}
			\caption{The MNL+Substitution.}
		\end{subfigure}
		\caption{SNR curves of the IM and the MNL, $OSR=4$, $LR=\frac{1}{2}$, $\lambda=2$.}
		\label{fig:1D_IM2}
	\end{figure}

\begin{figure}[tbh]
		\centering
		\begin{subfigure}[h]{.48\linewidth}
			\includegraphics[width=\textwidth,height=.6\textwidth]{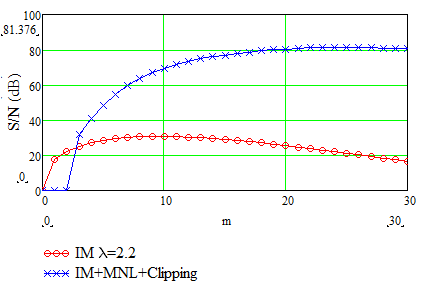}
			\caption{The MNL+Clipping.}
		\end{subfigure}
		\begin{subfigure}[h]{.48\linewidth}
			\includegraphics[width=\textwidth,height=.6\textwidth]{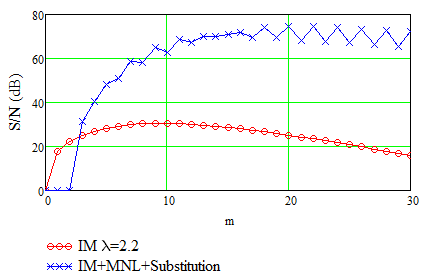}
			\caption{The MNL+Substitution.}
		\end{subfigure}
		\caption{SNR curves of the IM and the MNL, $OSR=8$, $LR=\frac{1}{3}$, $\lambda=2.2$.}
		\label{fig:1D_IM3}
	\end{figure}

\begin{figure}[htb]
		\centering
		\includegraphics[width=\textwidth,height=.3\textwidth]{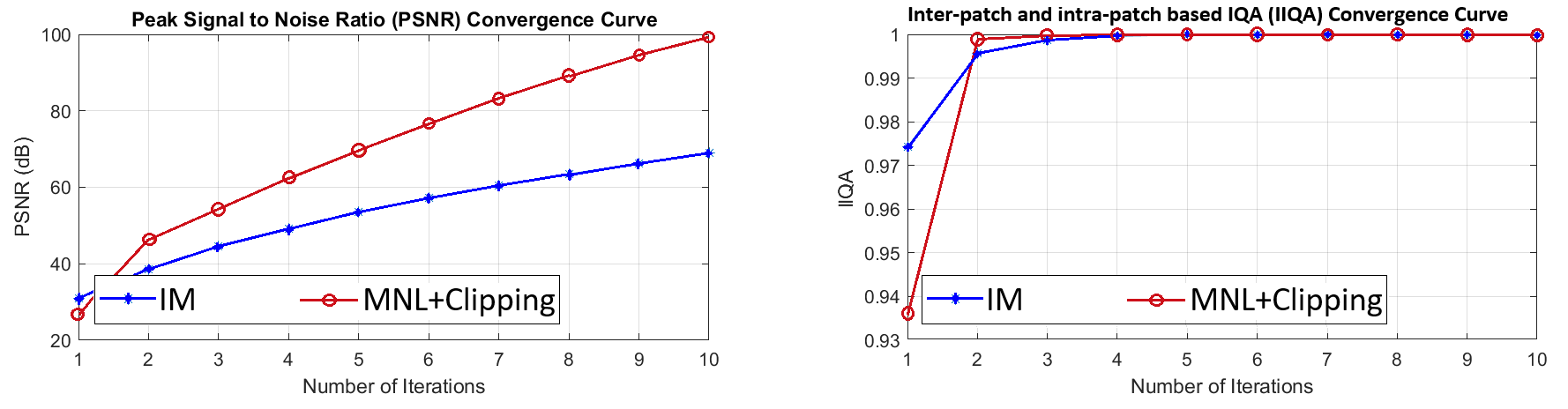}
		\caption{PSNR and IIQA curves of the IM and the MNL(+Clipping), $\lambda=2$, $OSR=4$, $LR=\frac{1}{3}$, image: Lenna.}
		\label{fig:2D_IM}
	\end{figure}

\begin{figure}[htb]
		\centering
		\includegraphics[width=\textwidth,height=.3\textwidth]{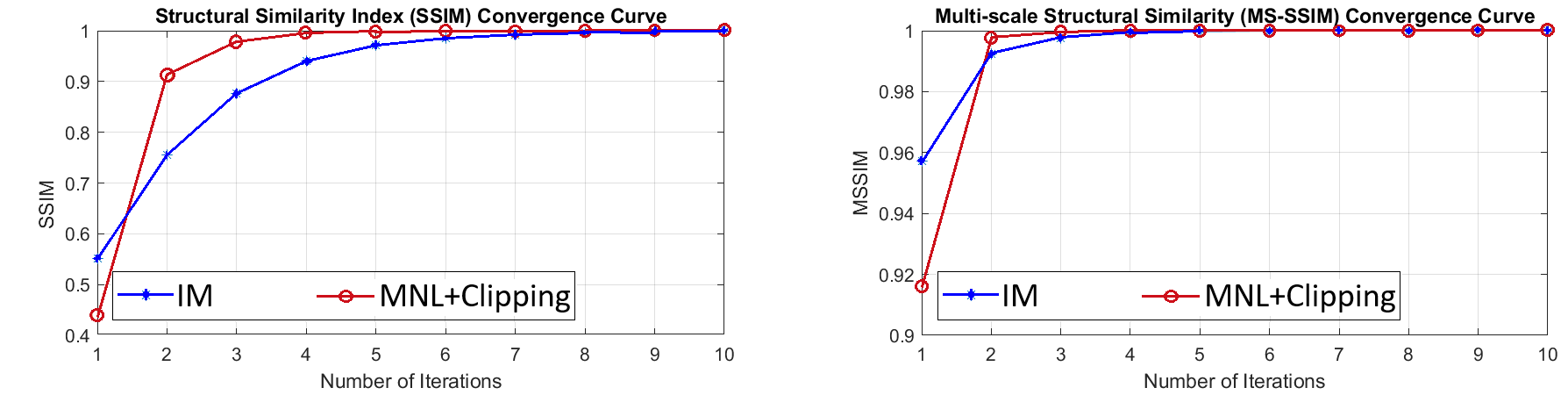}
		\caption{SSIM and MS-SSIM curves of the IM and the MNL(+Clipping), $\lambda=2$, $OSR=4$, $LR=\frac{1}{3}$, image: Lenna.}
		\label{fig:2D_IM2}
	\end{figure}

\begin{figure}[htb]
		\centering
		\includegraphics[width=\textwidth,height=.65\textwidth]{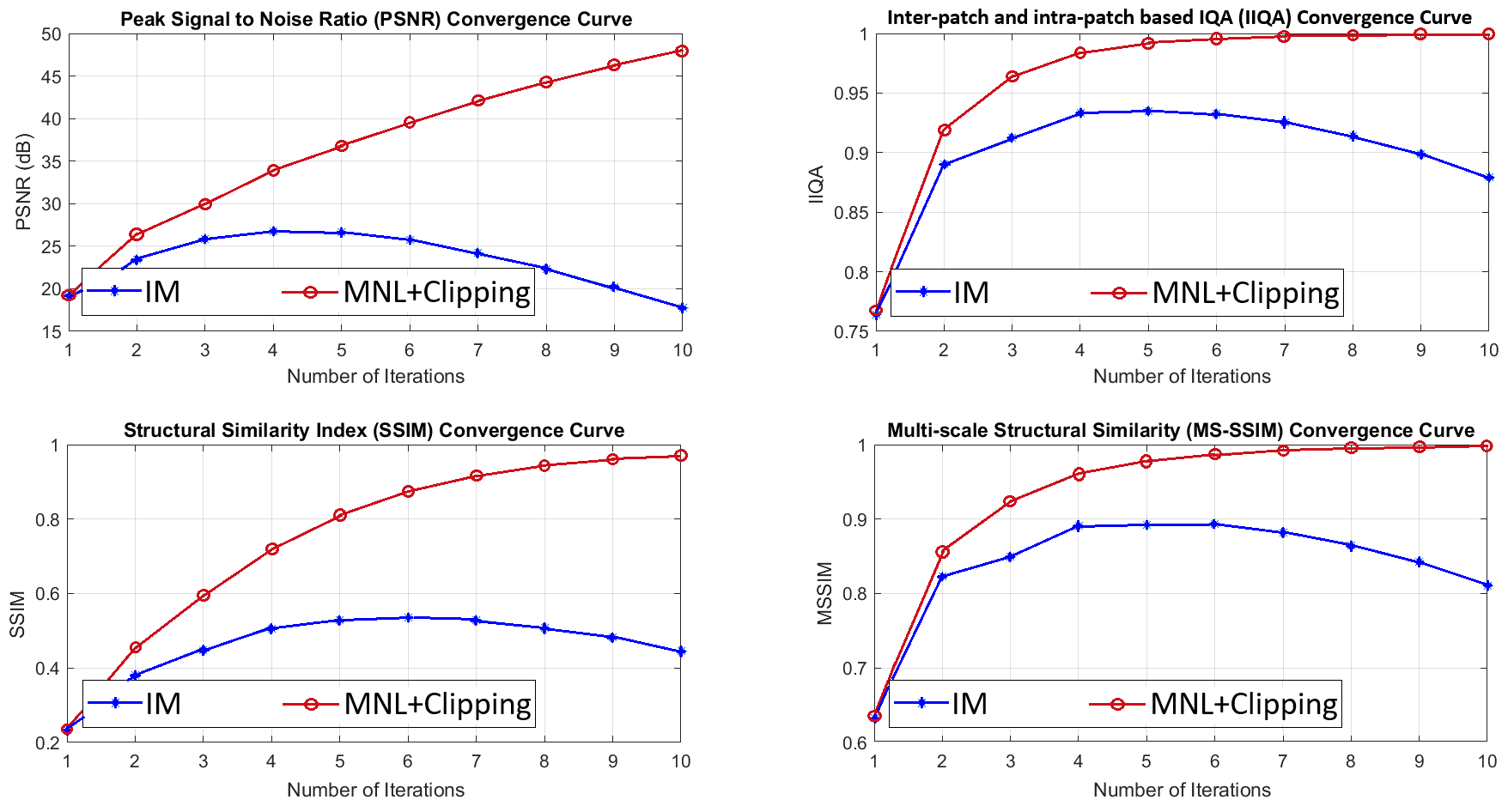}
		\caption{PSNR, IIQA, SSIM, and MS-SSIM curves of the IM and the MNL(+Clipping), $\lambda=3.5$, $OSR=4$, $LR=\frac{2}{3}$, image: Lenna.}
		\label{fig:2D_IM3}
	\end{figure}

\begin{figure}[htb]
		\centering
		\includegraphics[width=\textwidth,height=.3\textwidth]{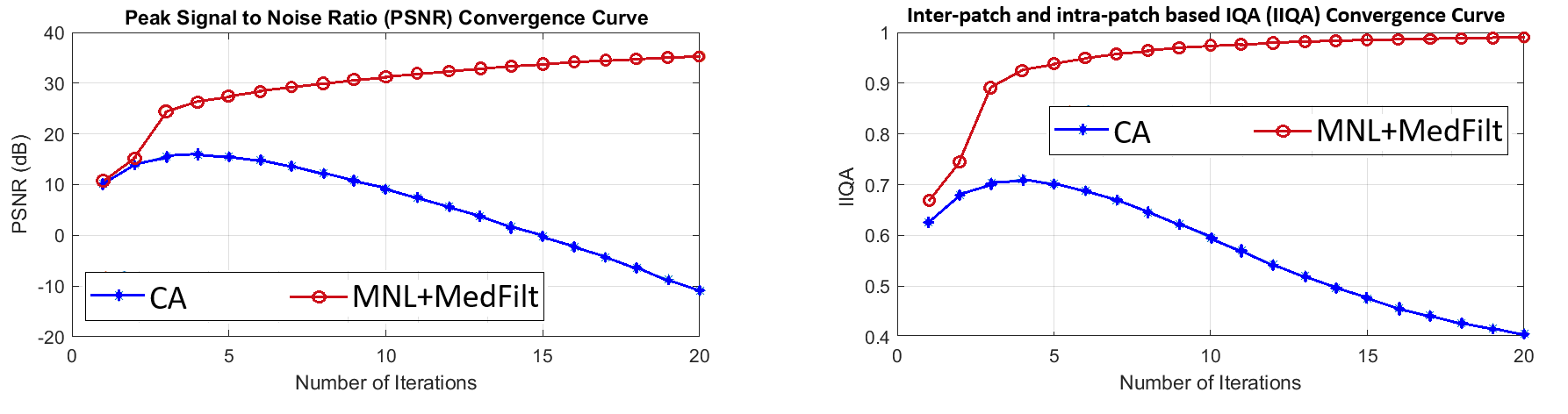}
		\caption{PSNR and IIQA curves of the IM and the MNL(+Clipping), $A=0.25$, $B=0.6$, $\lambda_0=3.5$, $LR=\frac{1}{2}$, image: Cameraman.}
		\label{fig:CA}
	\end{figure}

\begin{figure}[htb]
		\centering
		\includegraphics[width=\textwidth,height=.3\textwidth]{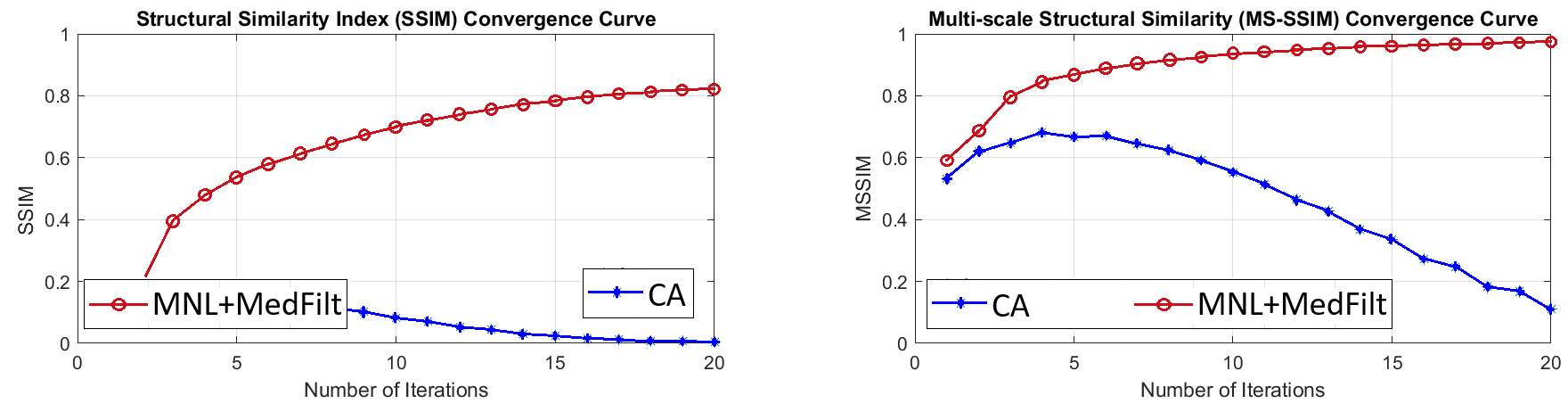}
		\caption{SSIM and MS-SSIM curves of the IM and the MNL(+Clipping), $A=0.25$, $B=0.6$, $\lambda_0=3.5$, $OSR=2$, $LR=\frac{1}{2}$, image: Cameraman.}
		\label{fig:CA2}
	\end{figure}

\begin{figure}[htb]
		\centering
		\begin{subfigure}[b]{.48\linewidth}
			\includegraphics[width=\textwidth,height=.9\textwidth]{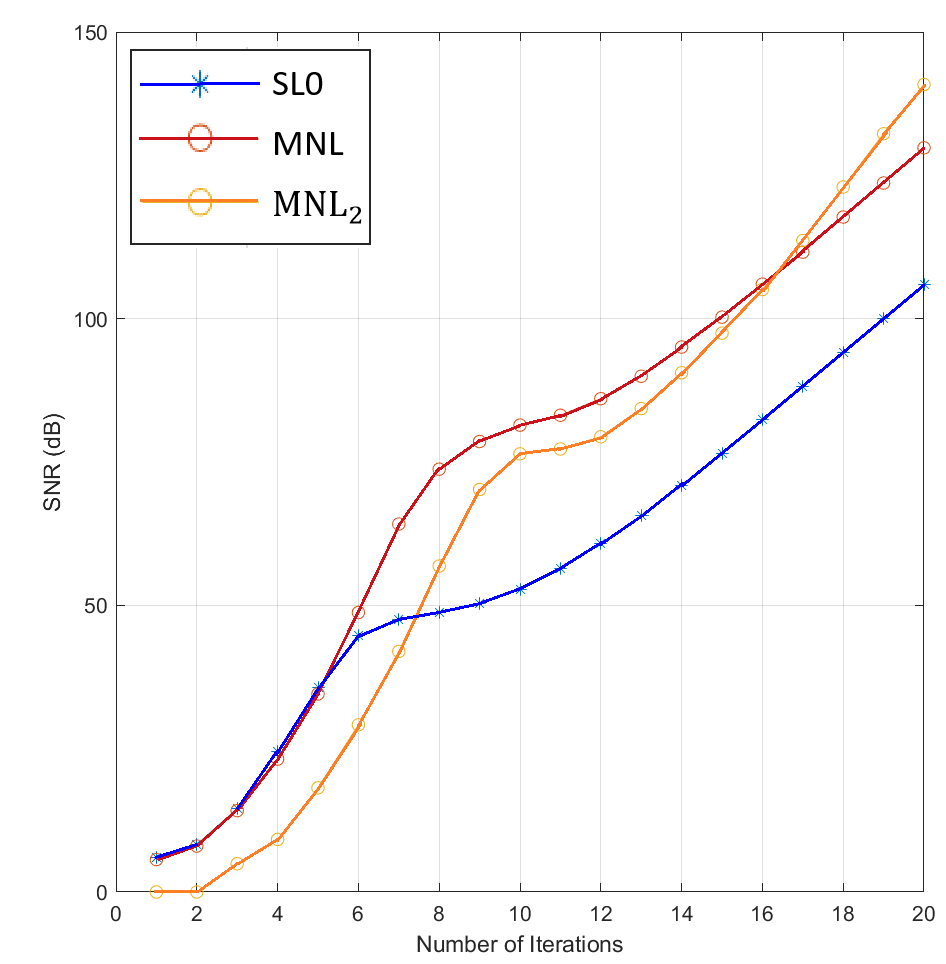}
			\caption{$LR=30\%$.}
		\end{subfigure}
		\begin{subfigure}[b]{.48\linewidth}
			\includegraphics[width=\textwidth,height=.9\textwidth]{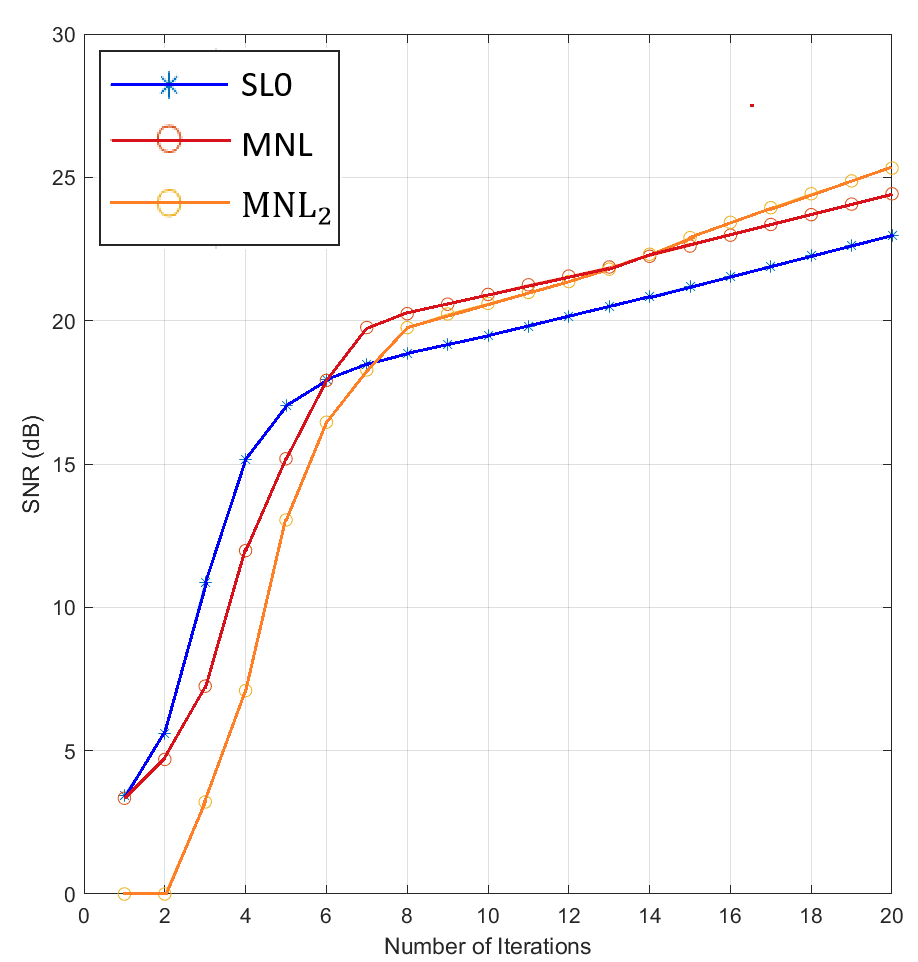}
			\caption{$LR=50\%$.}
		\end{subfigure}
		\caption{SNR curves of the SL0 recovery method and the MNL, $1D$ signal with $L=1000$, $SDF=0.5$, $p_{_{NZ}}=0.05$, averaged over $50$ runs.}
		\label{fig:sl0}
	\end{figure}

\begin{figure}[htb]
		\centering
		\begin{subfigure}[b]{.48\linewidth}
			\includegraphics[width=\textwidth,height=.9\textwidth]{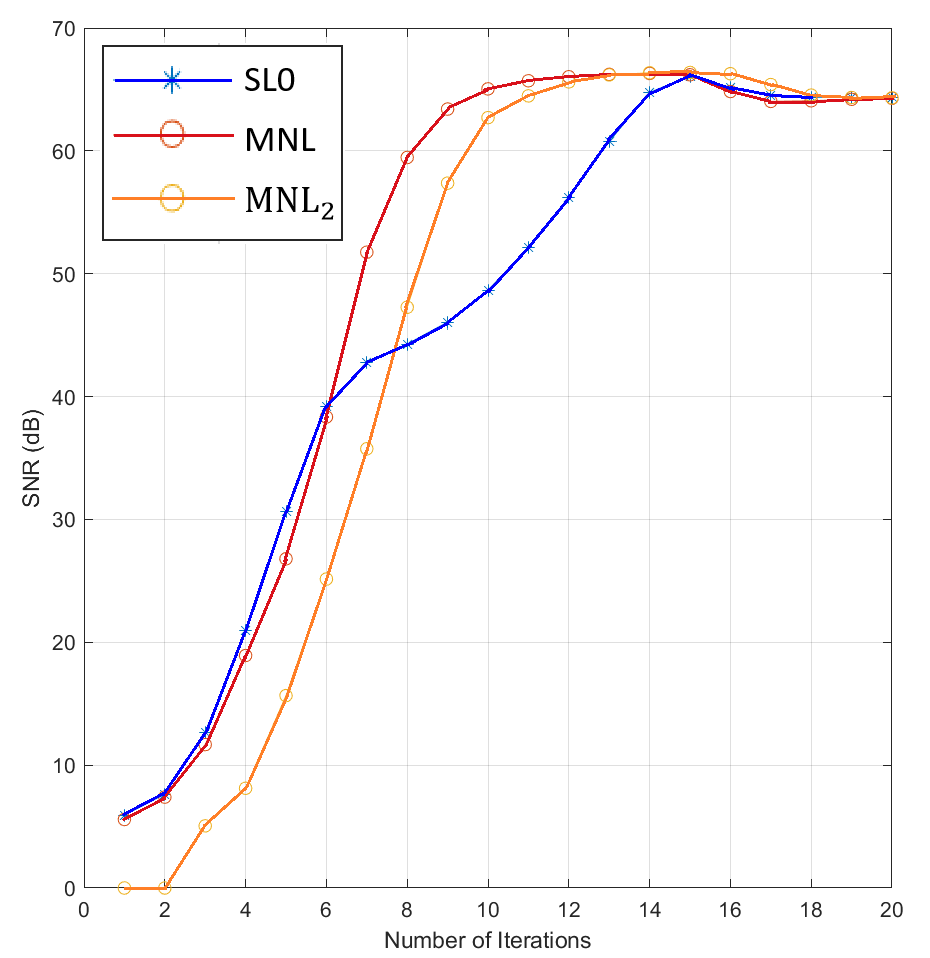}
			\caption{$SDF=0.5$.}
		\end{subfigure}
		\begin{subfigure}[b]{.48\linewidth}
			\includegraphics[width=\textwidth,height=.9\textwidth]{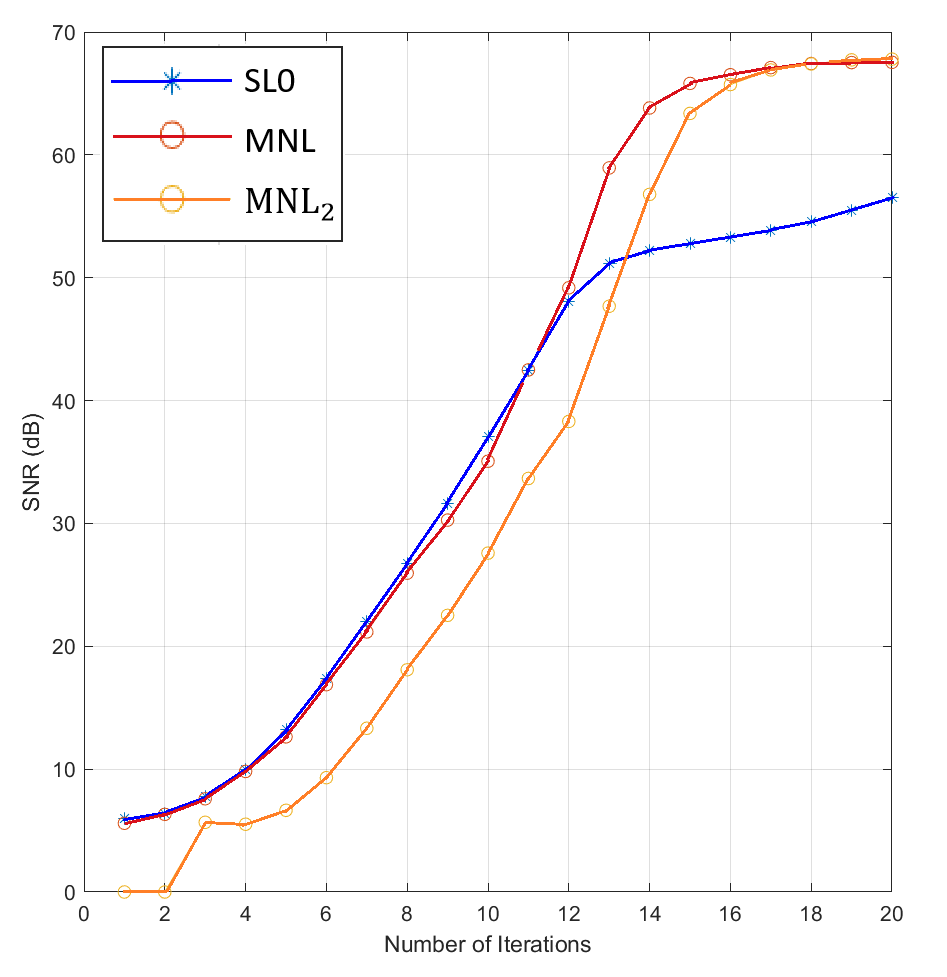}
			\caption{$SDF=0.7$.}
		\end{subfigure}
		\caption{SNR curves of the SL0 recovery method and the MNL, $1D$ signal with $L=1000$, $LR=30\%$, $p_{_{NZ}}=0.1$, averaged over $50$ runs.}
		\label{fig:sl0_2}
	\end{figure}

\begin{figure}[htb]
		\centering
		\includegraphics[width=\textwidth]{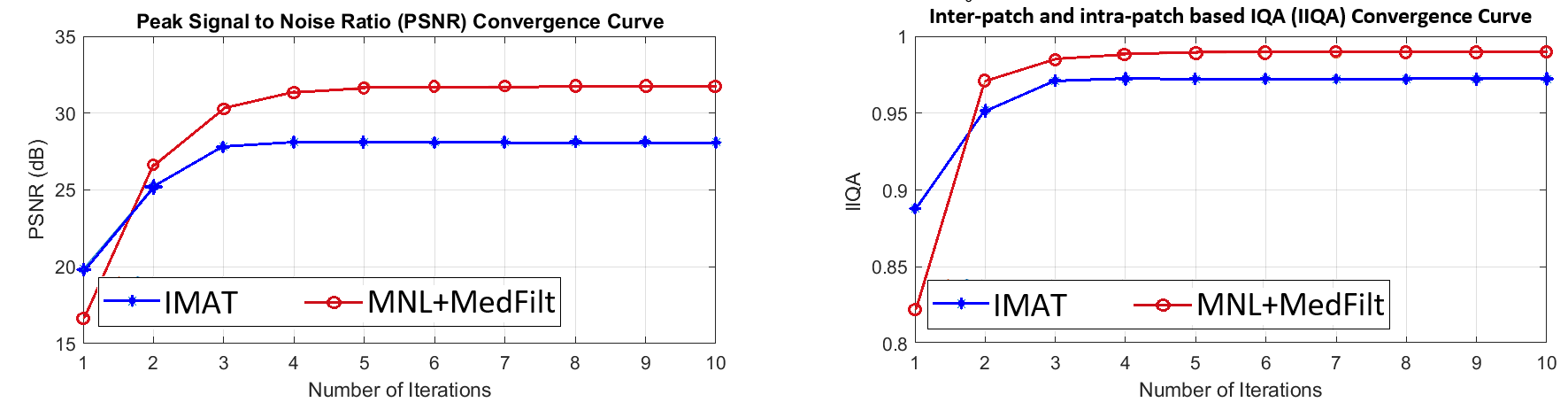}
		\caption{PSNR and IIQA curves of IMAT and the MNL(+MedFilt), $\lambda=2$, $T_0=300$, $\alpha=1$, $LR=30\%$, image: Baboon.}
		\label{fig:IMAT}
	\end{figure}

\begin{figure}[htb]
		\centering
		\includegraphics[width=\textwidth]{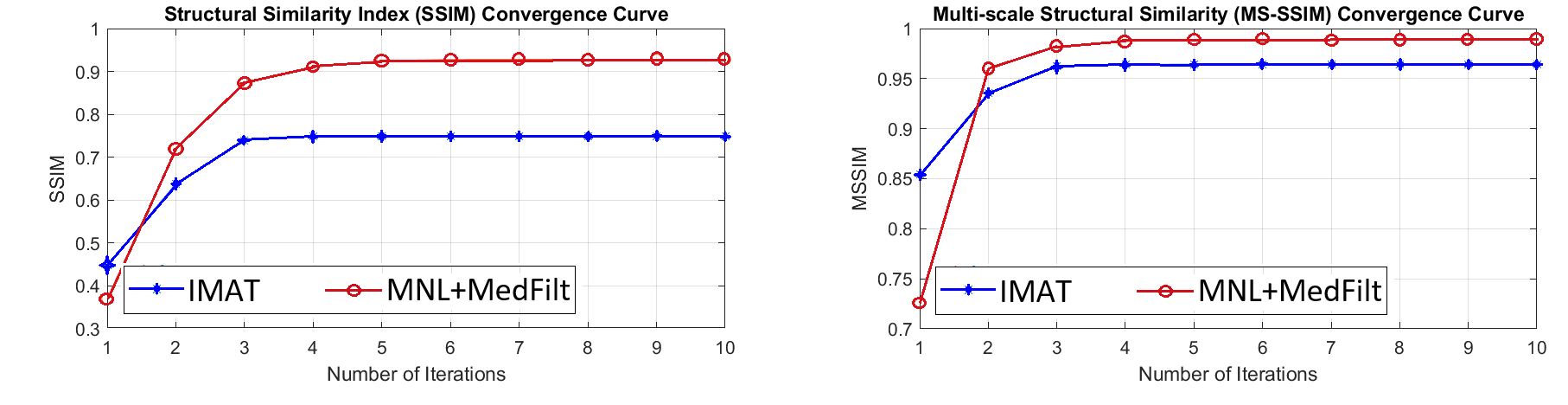}
		\caption{SSIM and MS-SSIM curves of IMAT and the MNL(+MedFilt), $\lambda=2$, $T_0=300$, $\alpha=1$, $LR=30\%$, image: Baboon.}
		\label{fig:IMAT2}
	\end{figure}

\begin{figure}[tbh]
		\centering
		\begin{subfigure}[h]{1\linewidth}
			\includegraphics[width=1\textwidth]{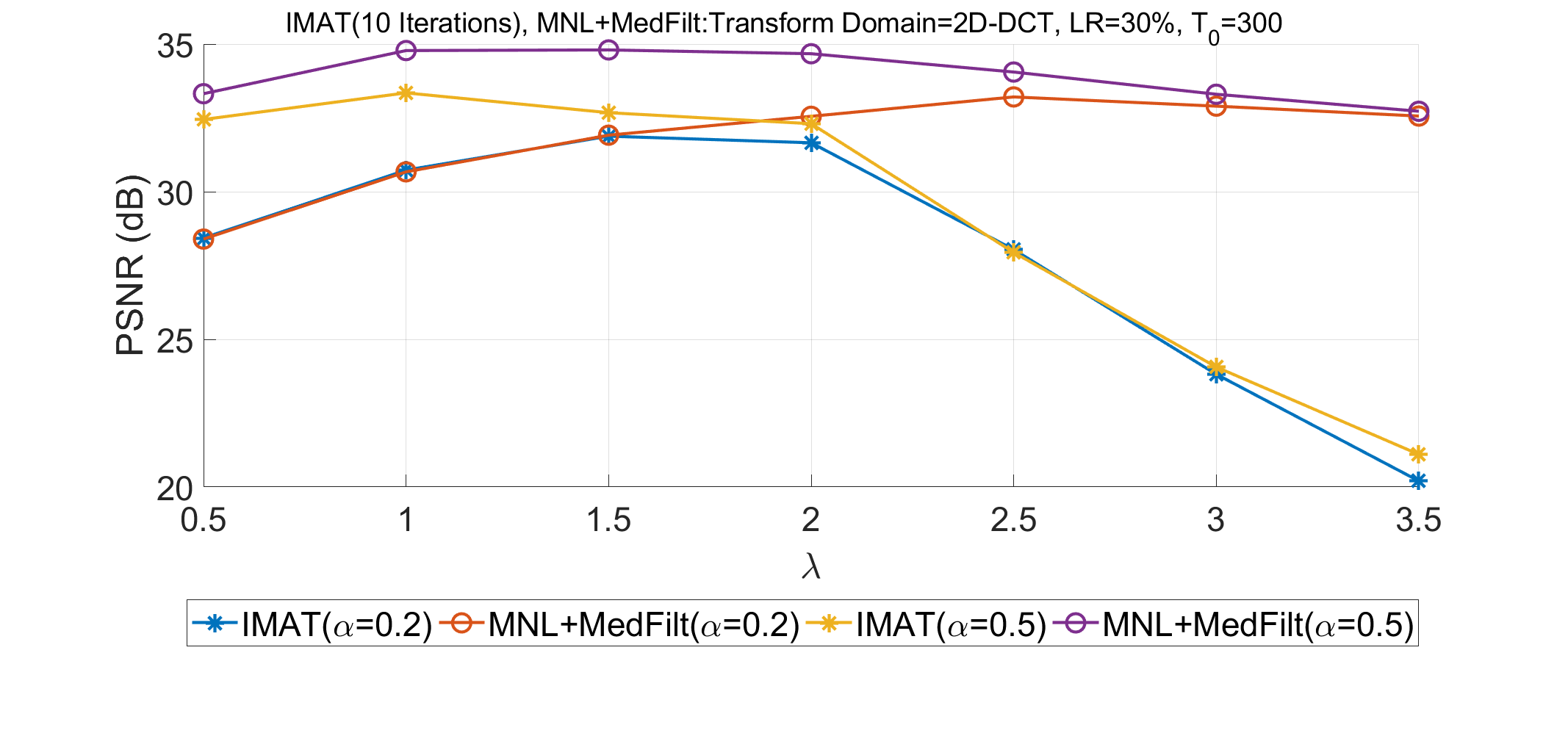}
			\caption{$LR=30\%$.}
		\end{subfigure}
		\begin{subfigure}[h]{1\linewidth}
			\includegraphics[width=1\textwidth]{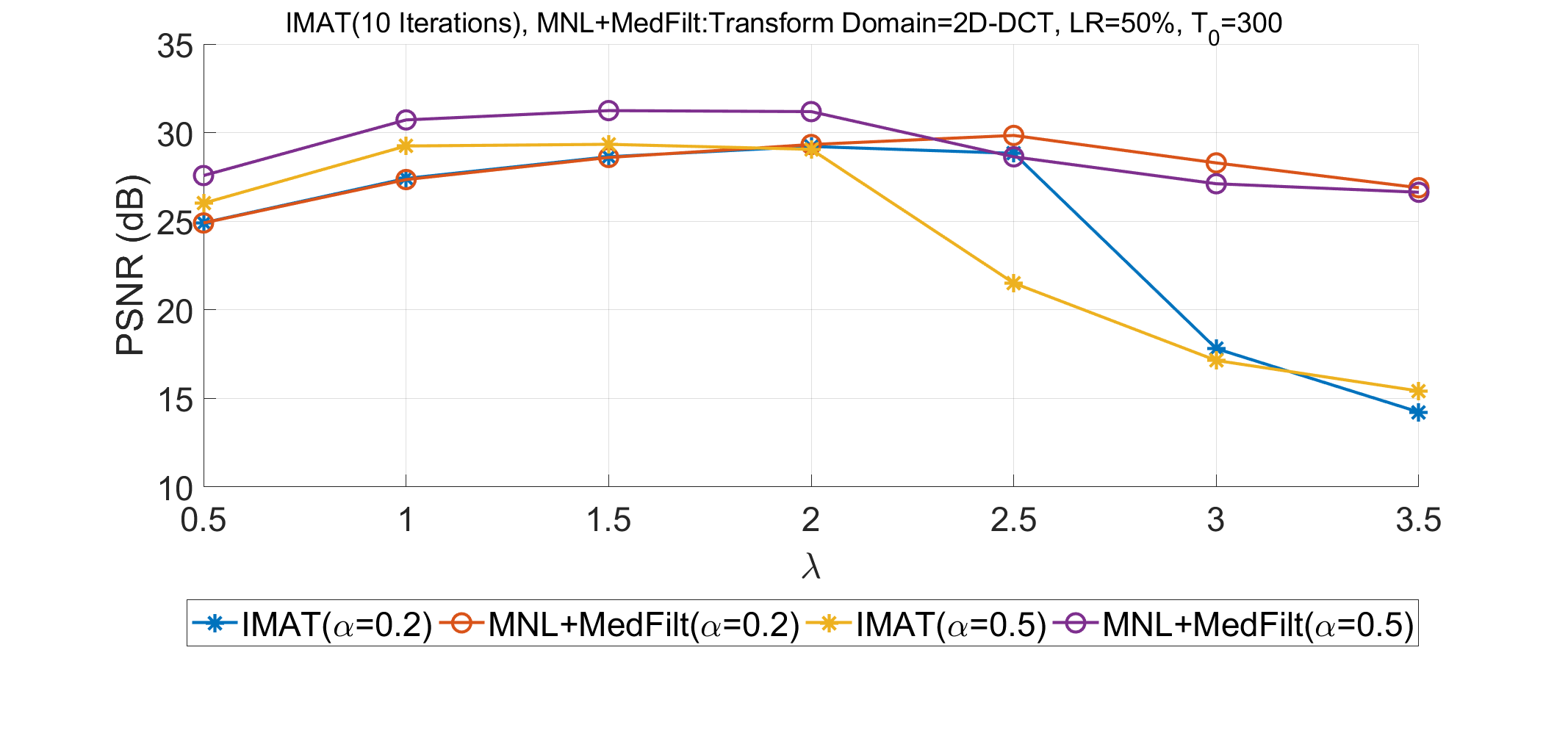}
			\caption{$LR=50\%$.}
		\end{subfigure}
		\caption{PSNR curves of the IMAT and the MNL (+MedFilt), image: Pirate.}
		\label{fig:IMAT3}
	\end{figure}

\begin{figure}[tbh]
		\centering
		\begin{subfigure}[h]{1\linewidth}
			\includegraphics[width=1\textwidth]{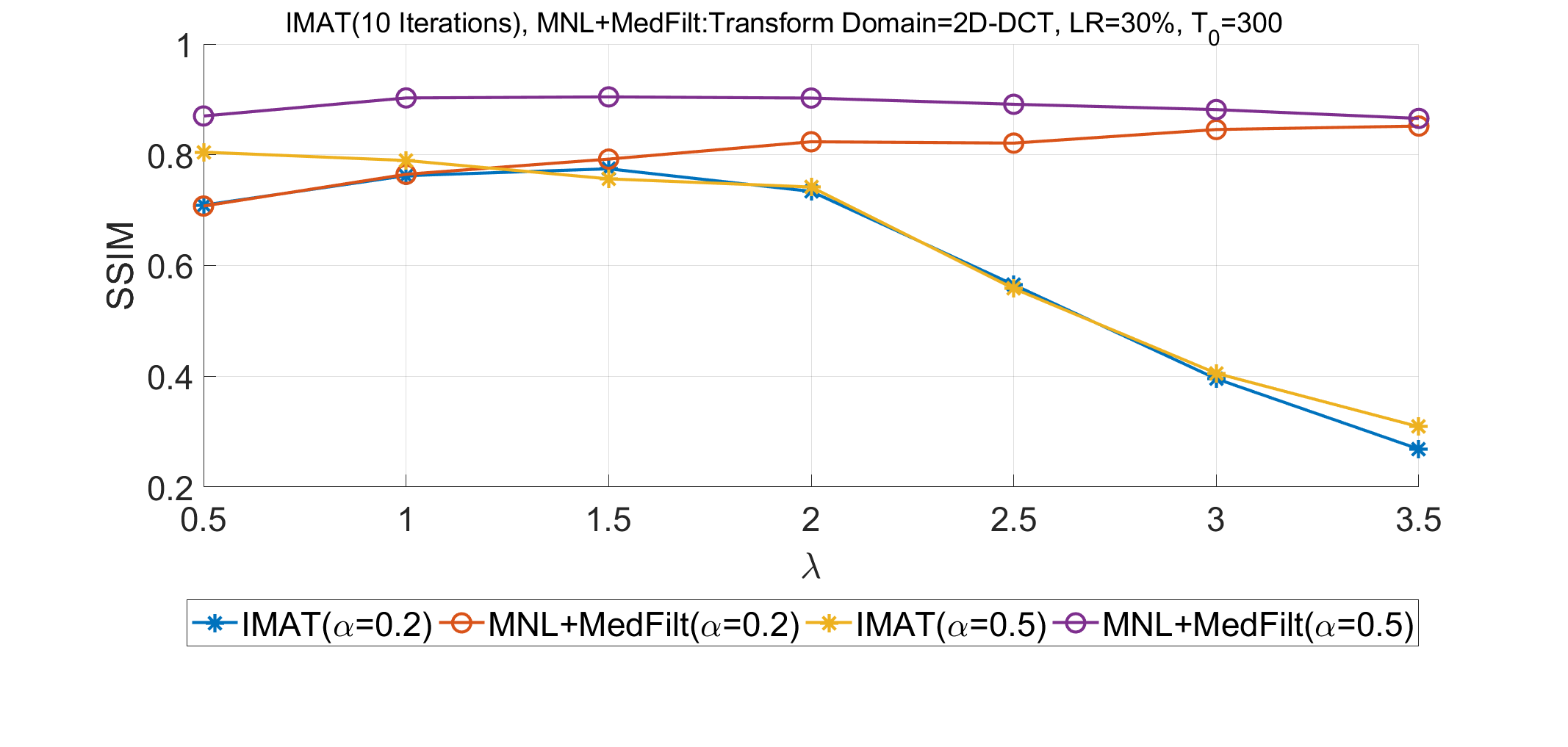}
			\caption{$LR=30\%$.}
		\end{subfigure}
		\begin{subfigure}[h]{1\linewidth}
			\includegraphics[width=1\textwidth]{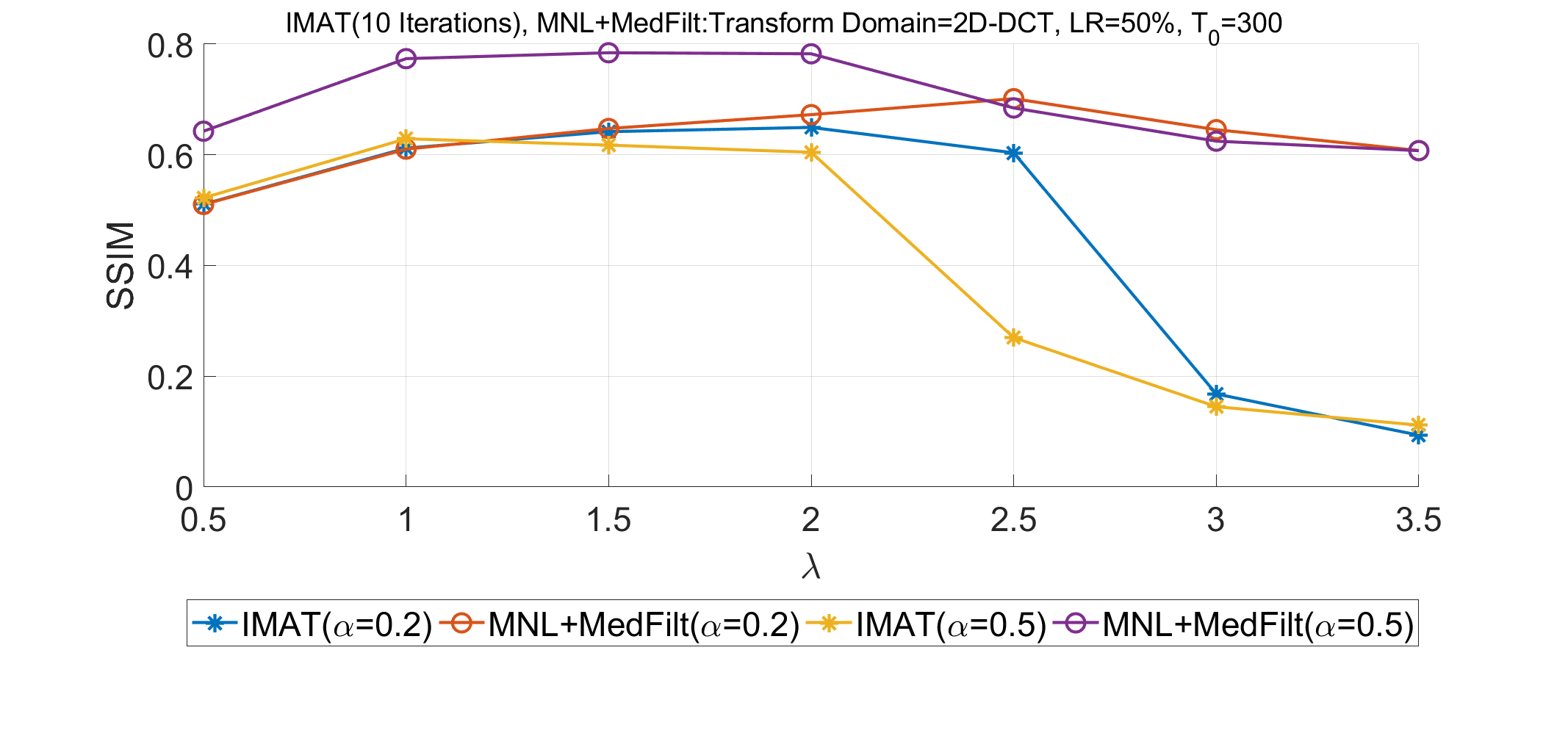}
			\caption{$LR=50\%$.}
		\end{subfigure}
		\caption{SSIM curves of the IMAT and the MNL (+MedFilt), image: Pirate.}
		\label{fig:IMAT4}
	\end{figure}

	For a selection of parameters same as the one in \cite{ghandi2008some} ($LR=33\%$ and $OSR=8$), the MNL can be stabilized by using Substitution and Clipping. Then, it can be applied to accelerate the IM even when $\lambda>1$, as shown in Fig.\ref{fig:1D_IM}.
	\newline
	By increasing and decreasing the LR and the OSR, respectively, the MNL starts to act unstably. This is because the chances for having a diverging case grow. Fortunately, simulation results show that the MNL estimated signal includes only a very few unstable points and therefore, can be stabilized using MedFilt and Substitution, as shown in Fig.\ref{fig:1D_IM2}. Note that the MNL improves iterative algorithms in terms of convergence. Hence, what actually leads to lower SNR improvement in this experiment is the performance of the IM.  
	\newline
	Increasing the relaxation parameter causes the IM to diverge. In order to avoid the latter problem, the MNL method can be used, as shown in Fig.\ref{fig:1D_IM3}. 
	\newline
	The MNL method can be easily generalized to improve iterative image recovery algorithms. It can be applied to the IM in order to reconstruct the image “Lenna” (with the size $512\times512$) from its nonuniform samples, as shown in Fig.\ref{fig:2D_IM}, \ref{fig:2D_IM2}.
	\newline
	The MNL was used to stabilize image recovery using the IM algorithm, as shown in Fig.\ref{fig:2D_IM3}.
	\newline
	Applying the MNL formula to the CA eventuates in the same results. Therefore, we focus on the stabilizing property of the MNL, as shown in Fig.\ref{fig:CA}, \ref{fig:CA2}.
	\newline
	There are two approaches to apply the MNL method to the SL0:
	\begin{enumerate}[I]
		\item	Applying the MNL formula to the last four estimations of the inner loop ($\text{MNL}$)
		\item 	Applying the MNL formula to the last estimations of the main algorithm ($\text{MNL}_2$)
	\end{enumerate}
	The signal is assumed to be sparse in the DFT domain.  Non-zero components are independently generated at random indices with the probability $p_{_{NZ}}$ while zero components are assumed to be contaminated by zero-mean Gaussian noise (with the standard deviation $\sigma_{off}$). The algorithm is initialized using the Least Square Estimation (LSE) and $\sigma_0=2\times\text{max}(|\hat{x}_0[n]|)$). $\sigma_{k}$ decreases by $SDF$ with each iteration. The inner minimization loop uses the SD method with 3 iterations and the adaption rate $\mu_0=2$. Fig.\ref{fig:sl0} and Fig.\ref{fig:sl0_2} show the results with and without the presence of noise, respectively.
	The MNL formula can be used to improve the performance of IMAT, as shown in Fig.\ref{fig:IMAT}, \ref{fig:IMAT2}. The latter figures show the performance curves of reconstructing the "Baboon" (also known as the "Mandrill") image from its nonuniform samples.
	\newline
	Considering IMAT's multiple parameters, it seems rather difficult to assess its sensitivity to parameter changes. To do so, the performance curves are depicted in terms of $\lambda$ and different values of  {\color{black} $\alpha$}. DCT was used as the function Tr(.) and the main algorithm was run in 5 iterations, as shown in Fig.\ref{fig:IMAT3},\ref{fig:IMAT4}.
	
	
\begin{figure}[tbh]
		\centering
		\begin{subfigure}[b]{1\linewidth}
			\includegraphics[width=\textwidth]{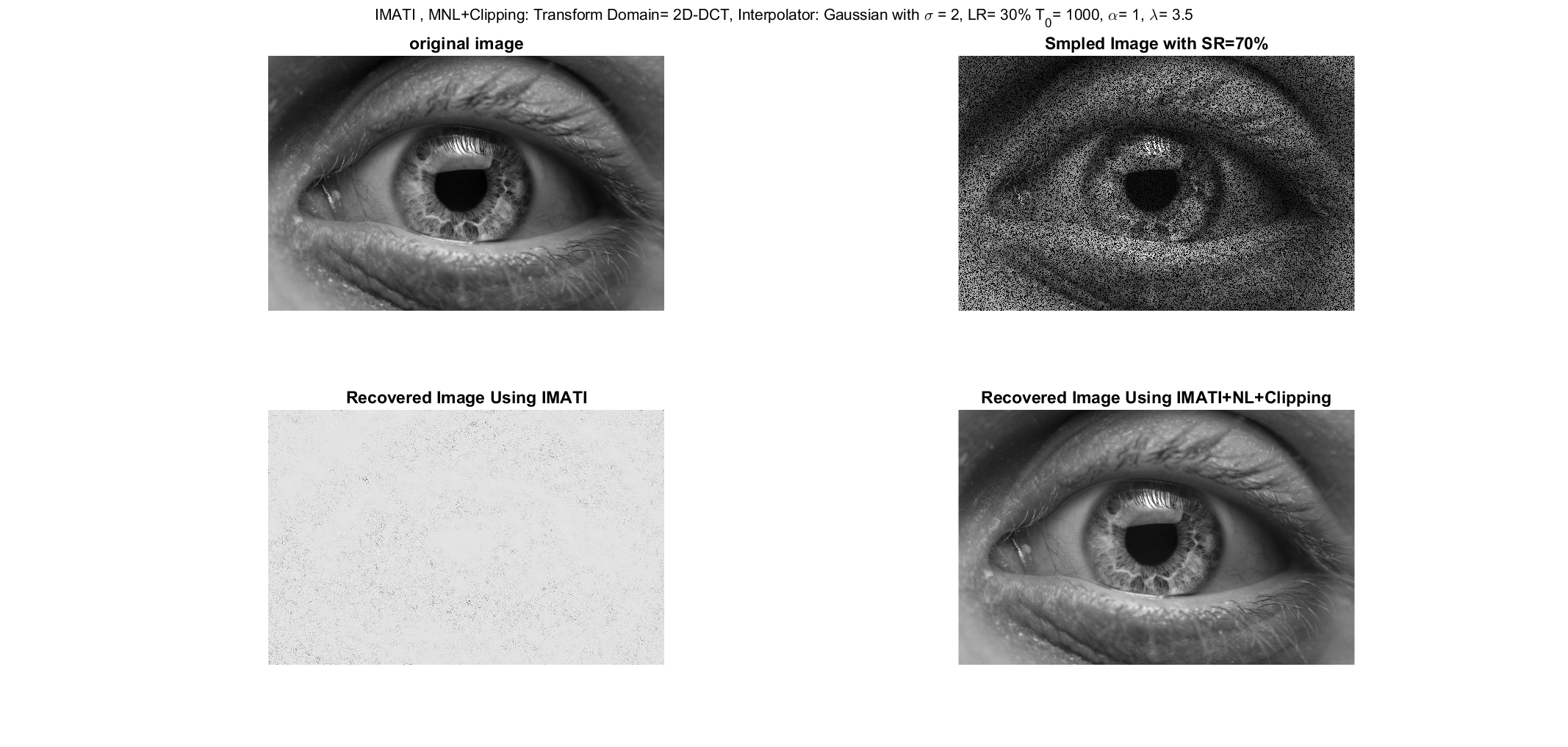}
			\caption{Main image, sampled image, and reconstructed images using IMATI and the MNL}
		\end{subfigure}
		\begin{subfigure}[b]{1\linewidth}
			\includegraphics[width=\textwidth]{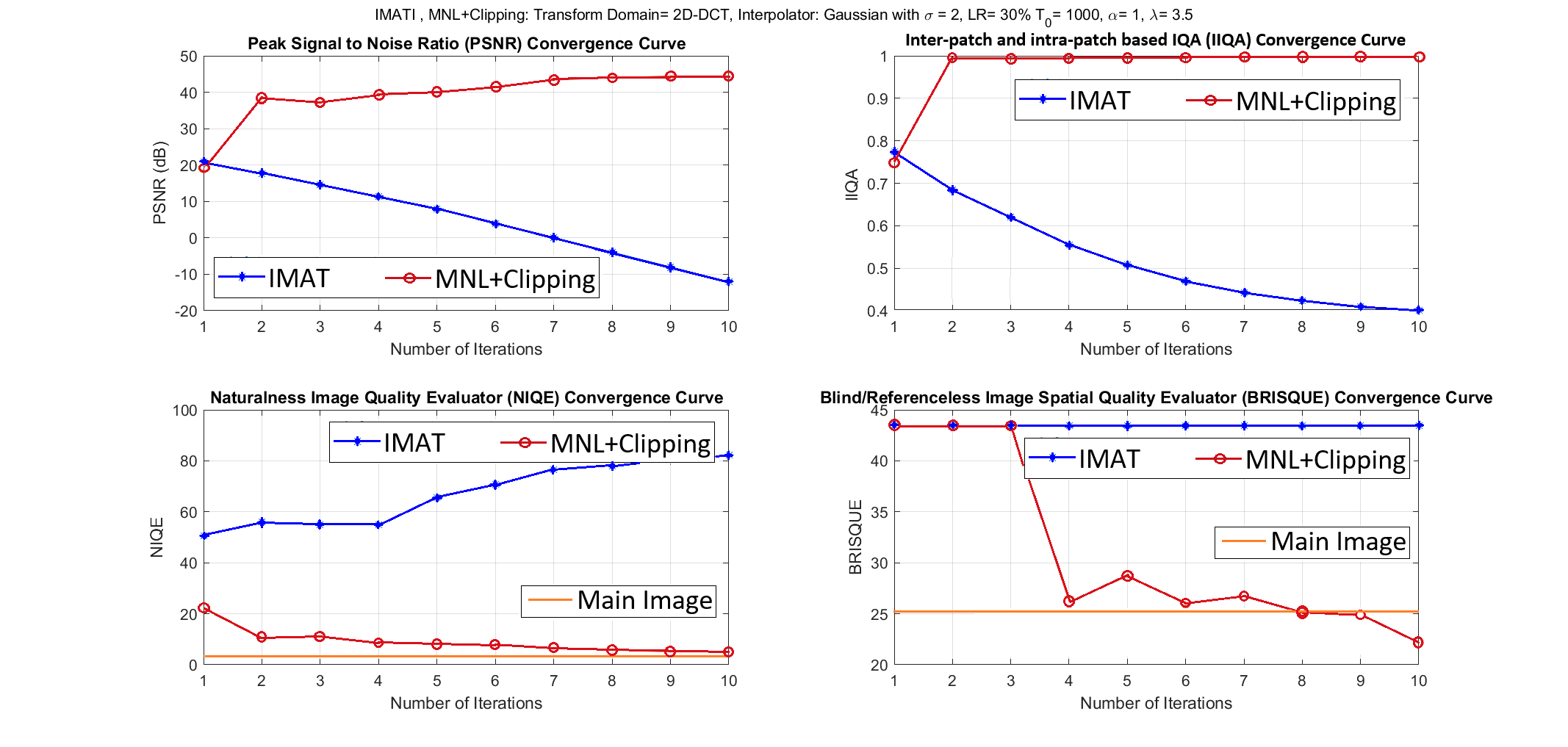}
			\caption{PSNR, IIQA, NIQE, and BRISQUE curves.}
		\end{subfigure}
		\begin{subfigure}[b]{1\linewidth}
			\includegraphics[width=\textwidth]{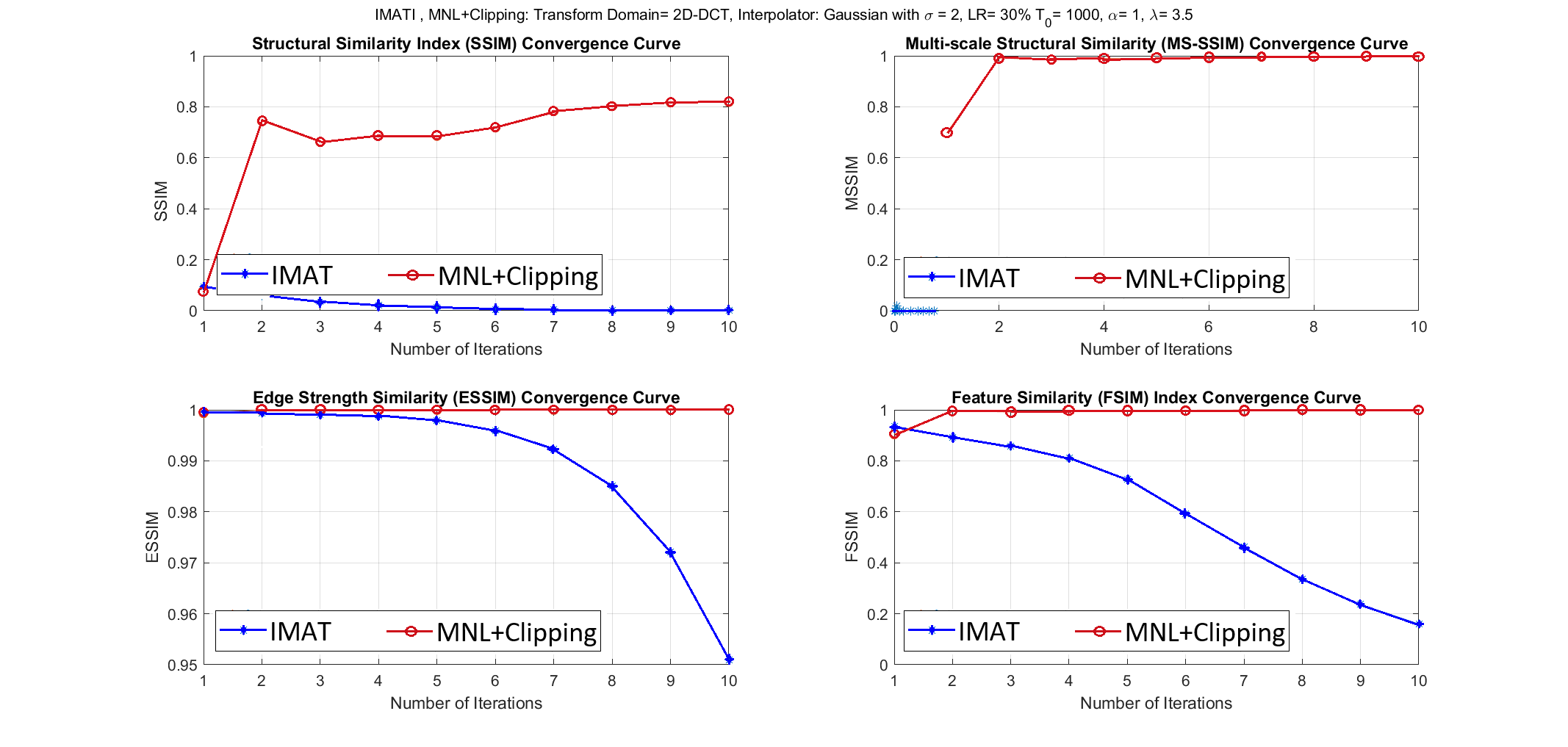}
			\caption{SSIM, MS-SSIM, ESSIM, and FSIM curves.}
		\end{subfigure}
		\caption{Image recovery using IMATI (+Gaussian interpretor) and the MNL (+Clipping), $\sigma=2$, $LR=30\%$, $T_0=1000$, $\alpha=1$, $\lambda=3.5$.}
		\label{fig:imati_eye}
	\end{figure}

\begin{figure}[tbh]
		\centering
		\begin{subfigure}[h]{1\linewidth}
			\includegraphics[width=1\textwidth]{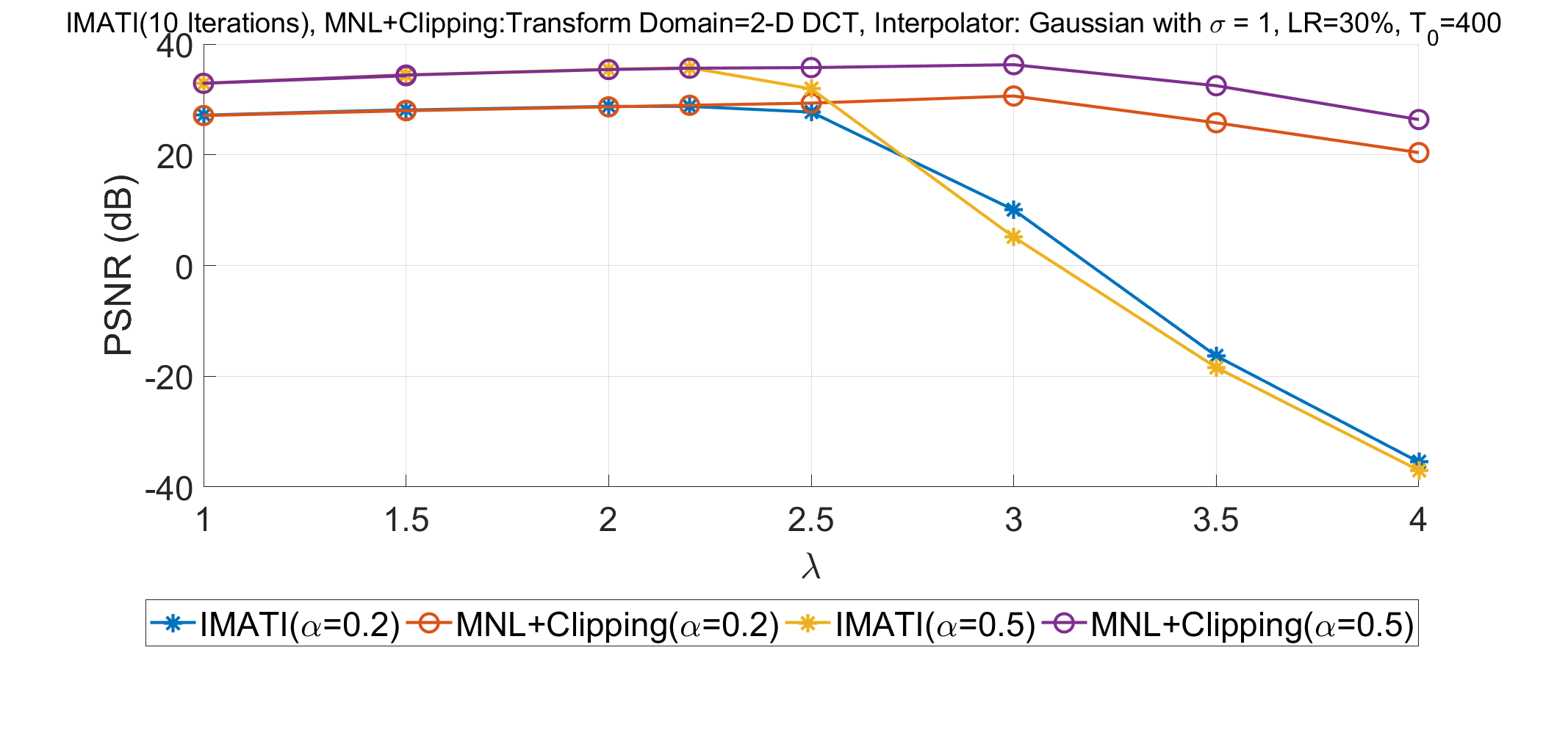}
			\caption{$LR=30\%$.}
		\end{subfigure}
		\begin{subfigure}[h]{1\linewidth}
			\includegraphics[width=1\textwidth]{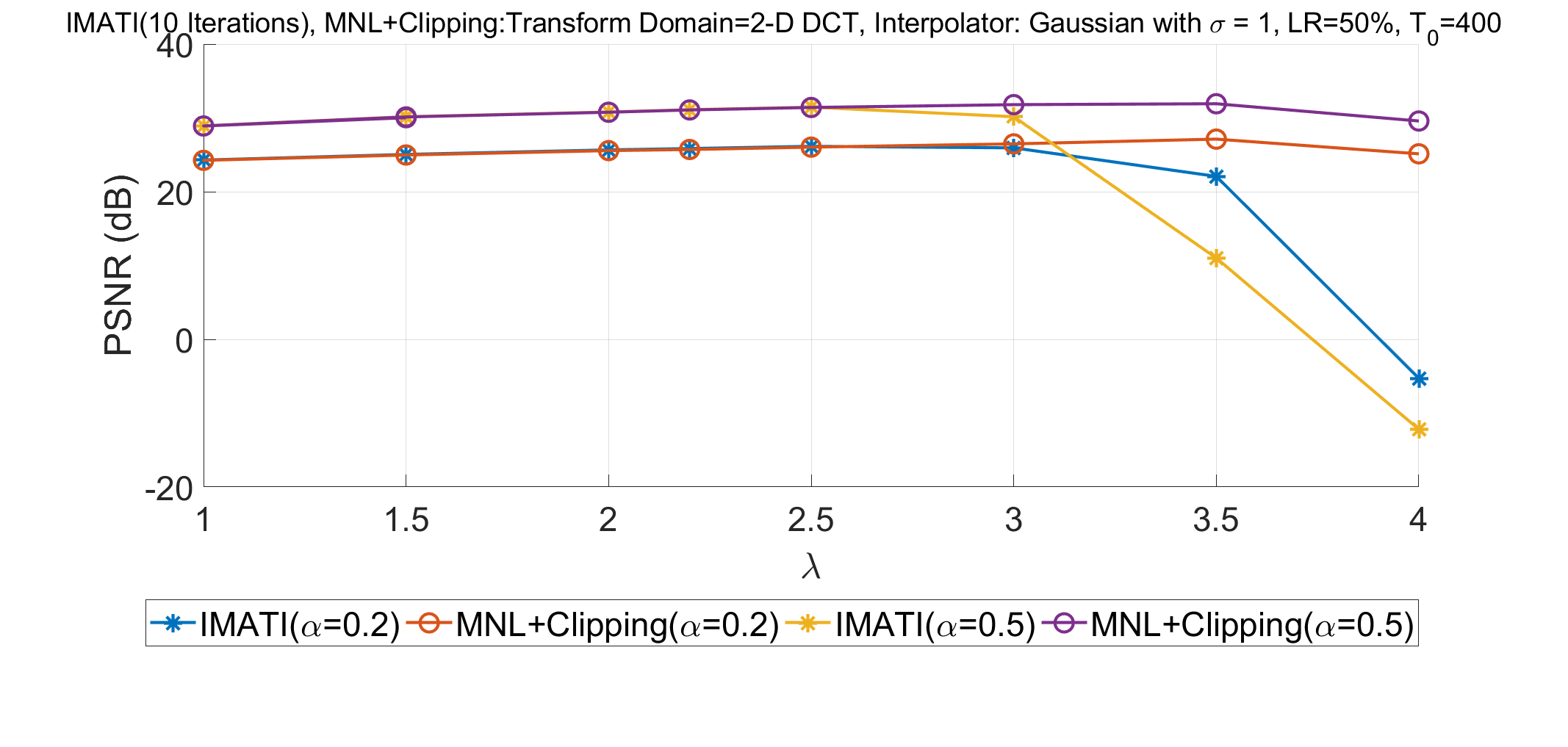}
			\caption{$LR=50\%$.}
		\end{subfigure}
		\caption{PSNR curves of IMATI and the MNL (+Clipping), image: Mandril.}
		\label{fig:IMATI}
	\end{figure}

\begin{figure}[tbh]
		\centering
		\begin{subfigure}[h]{1\linewidth}
			\includegraphics[width=1\textwidth]{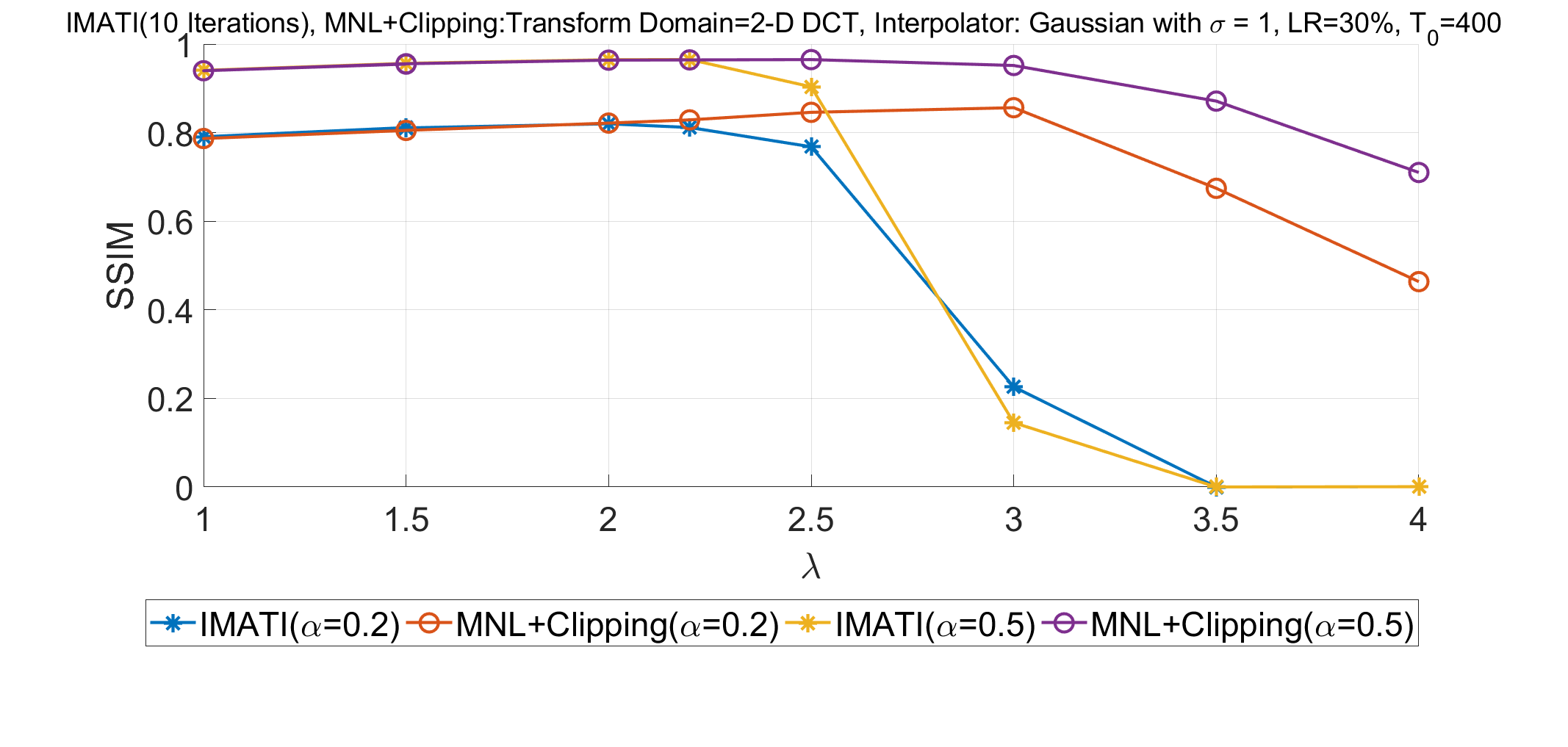}
			\caption{$LR=30\%$.}
		\end{subfigure}
		\begin{subfigure}[h]{1\linewidth}
			\includegraphics[width=1\textwidth]{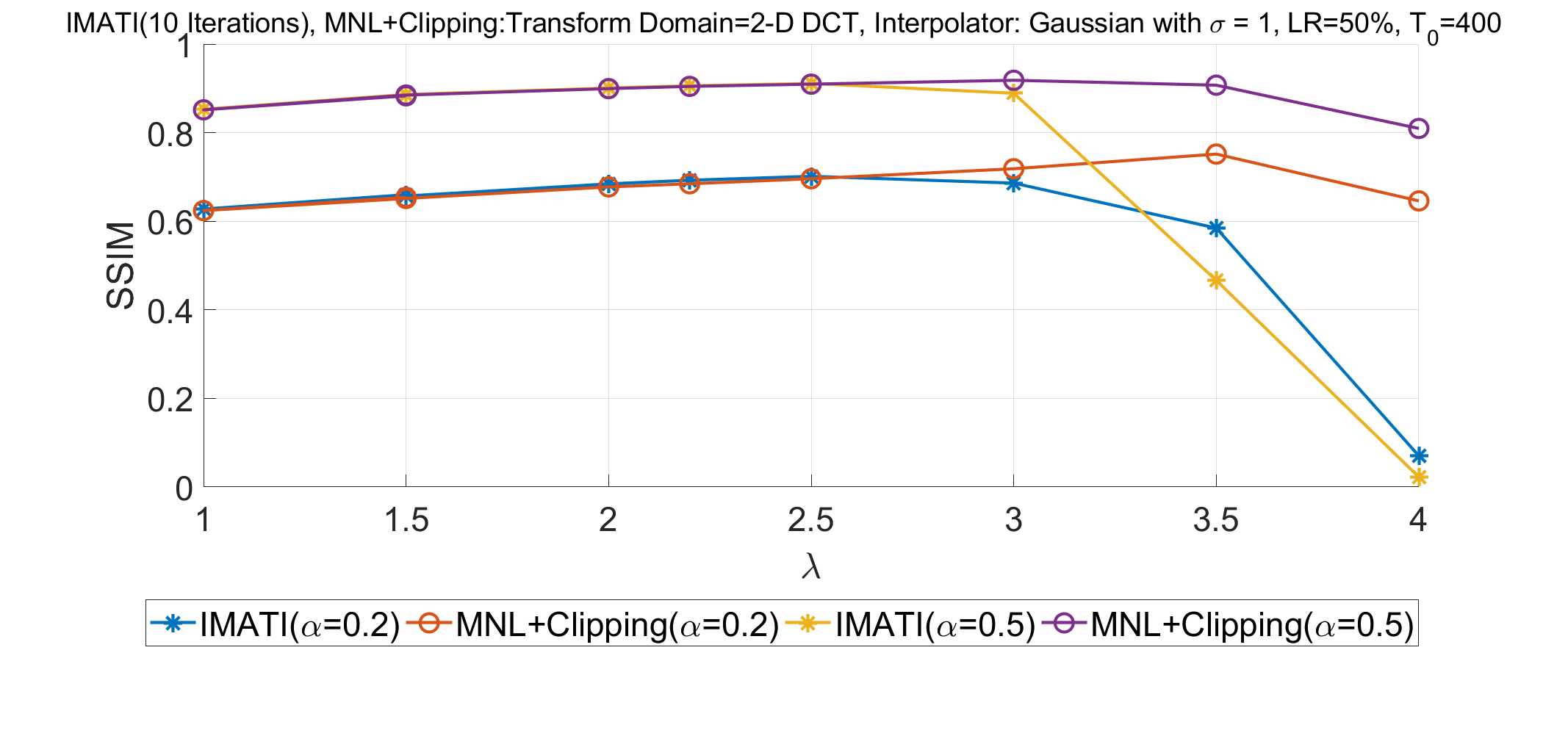}
			\caption{$LR=50\%$.}
		\end{subfigure}
		\caption{SSIM curves of IMATI and the MNL (+Clipping), image: Mandril.}
		\label{fig:IMATI2}
	\end{figure}

	As it can be seen, the MNL can preserve the performance of the algorithm to a great extent. Also, the stability range of the algorithm is extended. A wide range of stability is important because of its changes for different images and sampling patterns since in each case there is an optimum value of $\lambda$ for which the algorithm is both stable and fast enough. Crossing that optimal point causes the algorithm to diverge. Hence, by extending the stability range the reliability of the main algorithm increases. A highly detailed image of an eye (with the size $1600\times1030$) is reconstructed from its nonuniform samples as an example of sparse image recovery, as shown in Fig.\ref{fig:imati_eye}.
	\newline
	IMATI's sensitivity to parameter changes can be studied, as shown in Fig.\ref{fig:IMATI},\ref{fig:IMATI2}. 
	{\color{black} As shown in Fig. \ref{fig:lasso}, the MNL can be used to accelerate the	 ADMM algorithm for solving LASSO problems family with $\alpha, \rho, \lambda$ and $K$ as the over-relaxation, augmented Lagrangian, Lagrangian parameters and group size, respectively.
		\begin{figure}[tbh]
			\centering
			\begin{subfigure}[h]{1\linewidth}
				\includegraphics[width=1\textwidth]{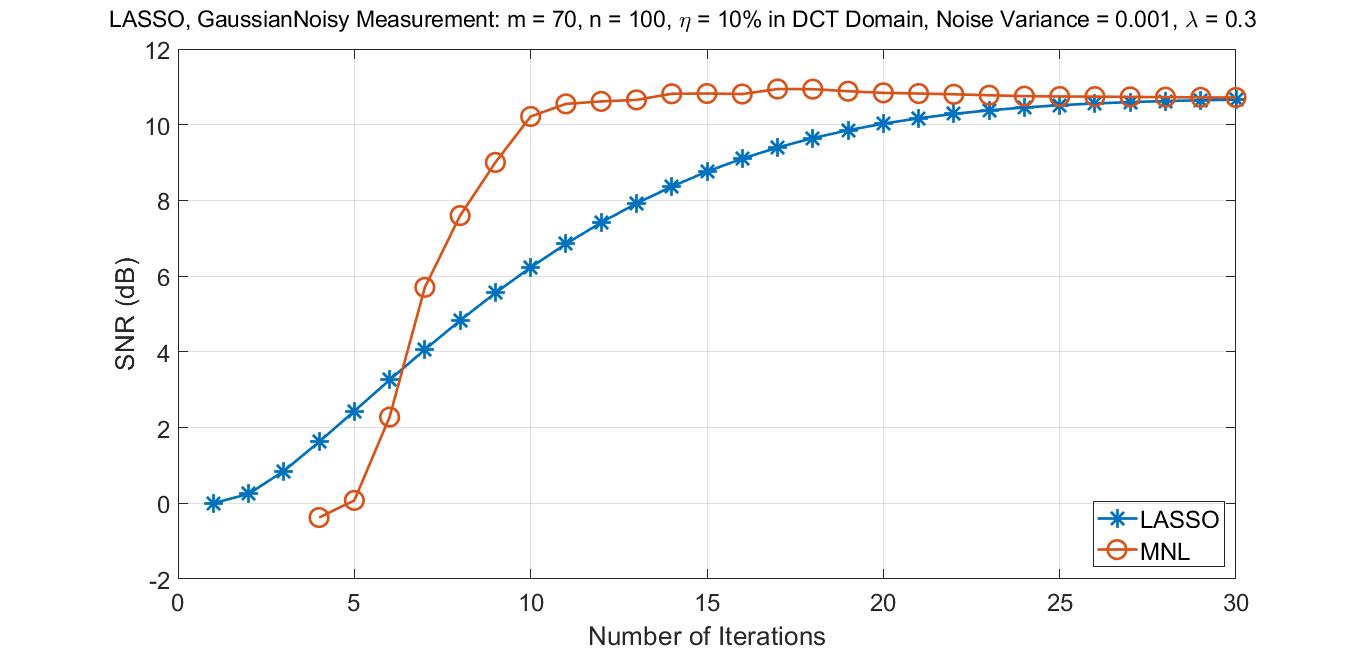}
				\caption{$\alpha=0.3, \rho=0.8$.}
			\end{subfigure}
			\begin{subfigure}[h]{1\linewidth}
				\includegraphics[width=1\textwidth]{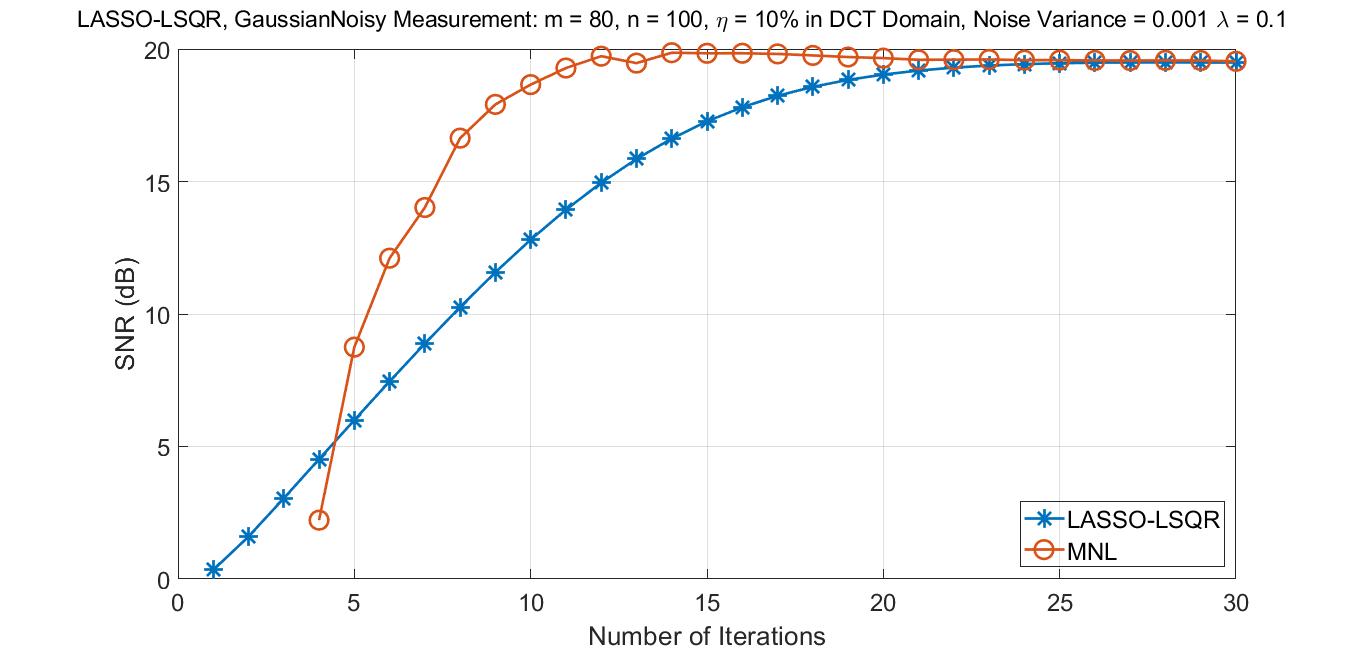}
				\caption{$\alpha=0.3, \rho=0.5$.}
			\end{subfigure}
			\begin{subfigure}[h]{1\linewidth}
				\includegraphics[width=1\textwidth]{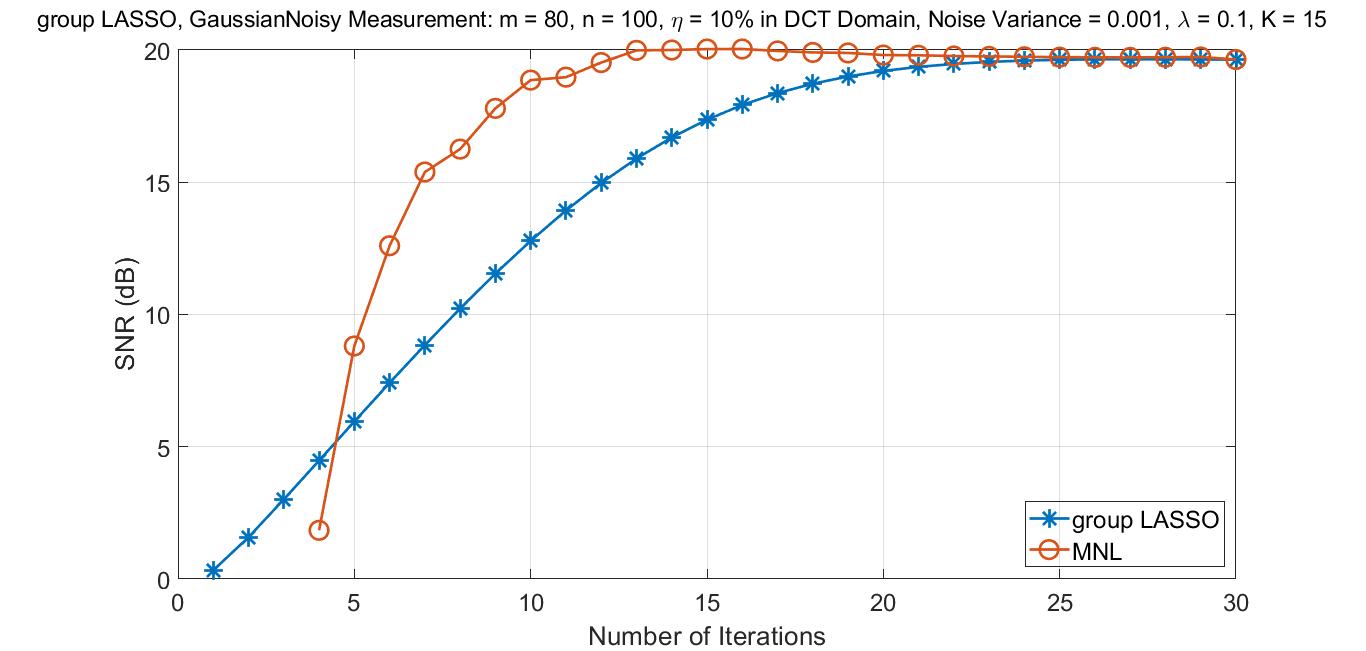}
				\caption{$\alpha=0.3, \rho=0.5$.}
			\end{subfigure}
			\caption{SNR curves of ADMM for solving LASSO, LASSO-LSQR and group LASSO, number of simulations = 100.}
			\label{fig:lasso}
		\end{figure}
		It must be mentioned that in the case of $\frac{m}{n}<0.5$, our simulations show that the MNL is not able to improve the ADMM; in fact, based on the Convergence Analysis, corresponding cases to $e_1\times e_3<0$ randomly occur and the NL diverges.
		\newline
		The convergence of the IRLS can be increased by MNL, as shown in Fig.\ref{fig:irls}.
		\begin{figure}[tbh]
			\centering
			\begin{subfigure}[h]{1\linewidth}
				\includegraphics[width=1\textwidth]{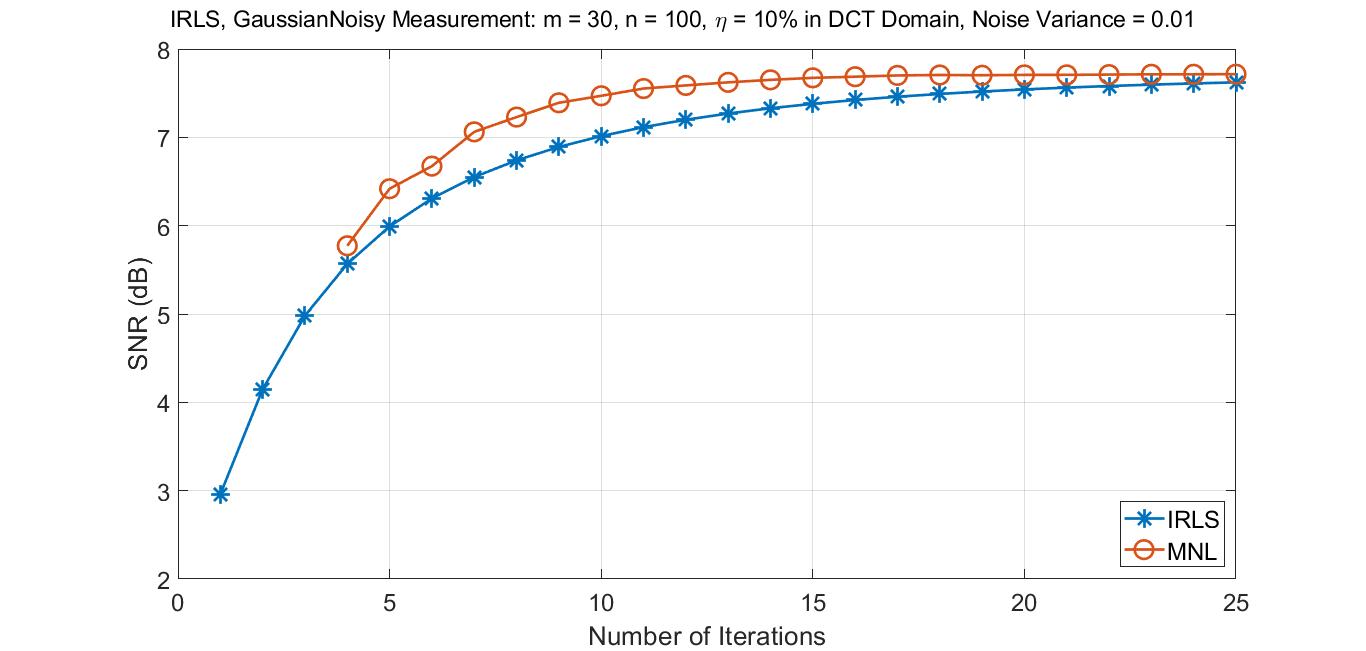}
			\end{subfigure}
			\begin{subfigure}[h]{1\linewidth}
				\includegraphics[width=1\textwidth]{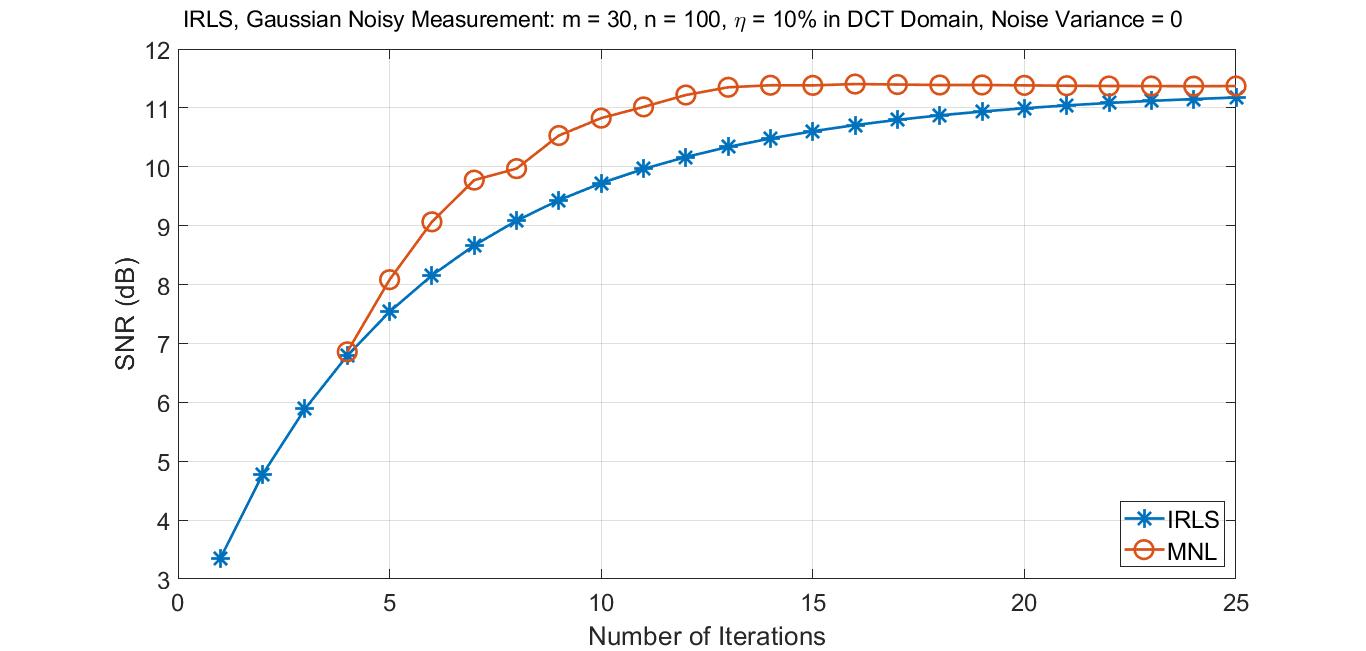}
			\end{subfigure}
			\caption{SNR curves of the IRLS, number of simulations = 100.}
			\label{fig:irls}
	\end{figure}}
	\section {Conclusion}
	\label{sec:conclusion}
	In this paper, a non-linear acceleration method (the NL) and its modification (the MNL) are introduced in order to accelerate and improve iterative algorithms. Besides, a complete analysis on convergence is given. It is stated that the proposed method can improve a wide variety of {\color{black}  linearly convergent algorithms} including optimization methods, band-limited signal recovery methods and sparse recovery algorithms. The proposed method is also capable of stabilizing diverging algorithms. It is shown that the MNL method improves iterative sparse recovery algorithms such as the IRLS, ADMM, SL0 and IMAT. Simulation results show that the performance of this method in terms of various quality assessments is noticeably better. By stabilizing and accelerating the CA method, it is shown that the MNL method can even be used to improve an acceleration method.
	\FloatBarrier
	\section*{Acknowledgement}
	{\color{black} We wish} to thank Nasim Bagheri for her assistance with the text editing.
	
	\appendix
	\section{Convergence Analysis}
	At first, assume that $|\alpha_1|=|\alpha_2|$ (linear convergence):
	\newline Cases $1.1$ and $1.2$, if $\alpha_1=\alpha_2=\pm\alpha$ ($\alpha>0$) then
	\begin{align}
	e_{_{NL}}=\frac{(\pm\alpha^{-1})(\pm\alpha)\times {e}_2^2-{e}_2^2}
	{(\pm\alpha^{-1}+(\pm\alpha)-2){e}_2}=\frac{(1-1)}{\pm\alpha^{-1}+(\pm\alpha)-2}e_2\notag
	\end{align}
	which is equal to zero as expected.
	\newline
	Case 1.3, if $\alpha_1=-\alpha_2=\alpha>0$ then
	\begin{align}
	e_{_{NL}}=\frac{(\alpha^{-1})(-\alpha)-1}
	{\alpha^{-1}-\alpha-2}{e}_2=\frac{-2}{\alpha^{-1}-\alpha-2}{e}_2\notag\\
	\Rightarrow |\frac{e_{_{NL}}}{e_3}|=\frac{2}{|\alpha^2+2\alpha-1|}\notag
	=\frac{2}{|(\alpha+1)^2-2|}
	\end{align}
	which means that for a convergent sequence ($0<\alpha<1$), the NL estimation diverges; in other words, we have $|\frac{e_{_{NL}}}{e_3}|>1$. 
	\newline
	Case 1.4, if $-\alpha_1=\alpha_2=\alpha>0$ then
	\begin{align}
	e_{_{NL}}=\frac{(-\alpha^{-1})(\alpha)-1}
	{-\alpha^{-1}+\alpha-2}{e}_2=\frac{-2}{-\alpha^{-1}+\alpha-2}{e}_2\notag\\
	\Rightarrow |\frac{e_{_{NL}}}{e_3}|=\frac{2}{|\alpha^2-2\alpha-1|}\notag
	=\frac{2}{|(\alpha-1)^2-2|}
	\end{align}
	which also results in the divergence of the NL method in the case of a converging sequence ($0<\alpha<1$). 
	
	In a more general case, it can be assumed that $|\alpha_1|\neq|\alpha_2|$. Hence, $e_1$ and $e_3$ should be rewritten in terms of $e_2$. In a converging algorithm, subsequent estimations satisfy
	\begin{equation}
	|e_1|>|e_2|>|e_3|.\notag
	\end{equation}
	We can assume that after each iteration, the algorithm becomes less capable of reducing the errors which results in 
	\begin{align}
	|\alpha_1|=\alpha\:,\:|\alpha_2|=(1+\delta)\alpha\quad;\quad\delta>0
	\label{eq:nl_1}.
	\end{align}
	The latter alongside the convergence of the algorithm leads to obtaining
	\begin{align}
	|\alpha_1|,|\alpha_2|<1\Rightarrow\alpha<1\:
	,\:\delta<\frac{1}{\alpha}-1.
	\label{eq:nl_2}
	\end{align}
	Hence, we have the following, which represent the case of sub-linear convergence:
	\newline
	Case 2.1, if all of the $e_i$'s have the same sign, in other words, if  $e_1\times e_2>0$ and $e_3\times e_2>0$, then
	\begin{align}
	e_{_{NL}}=\frac{\alpha^{-1}\times(1+\delta)\alpha-1}{\alpha^{-1}+(1+\delta)\alpha-2}e_2=\frac{\delta\alpha}{(\alpha-1)^2+\delta\alpha^2}\notag\\
	\Rightarrow|\frac{e_{_{NL}}}{e_3}|=\frac{\delta}{(1+\delta)((\alpha-1)^2+\delta\alpha^2)}\leq1\notag
	\end{align}
	where the equality holds for $\delta=\frac{1}{\alpha}-1$; this implies that the NL method does not improve the existing estimation, as expected from (\ref{eq:nl_1}) and (\ref{eq:nl_2}).
	\newline
	Case 2.2, if $e_1\times e_2<0$ and $e_3\times e_2<0$ then
	\begin{align}
	e_{_{NL}}=\frac{-\alpha^{-1}\times-(1+\delta)\alpha-1}{-\alpha^{-1}-(1+\delta)\alpha-2}e_2=\frac{\delta\alpha}{-(\alpha+1)^2-\delta\alpha^2}\notag\\
	\Rightarrow|\frac{e_{_{NL}}}{e_3}|=\frac{\delta}{(1+\delta)((\alpha+1)^2+\delta\alpha^2)}<1\:;\:\forall \delta>0.\notag
	\end{align}
	\newline
	Case 2.3, if $e_1\times e_2>0,e_3\times e_2<0$ and $0<\delta<\frac{1}{\alpha}-1$ then
	\begin{align}
	e_{_{NL}}=\frac{-\alpha^{-1}\times(1+\delta)\alpha-1}{\alpha^{-1}-(1+\delta)\alpha-2}e_2=
	\frac{-(\delta+2)\alpha}{-((\alpha+1)^2-2+\delta\alpha^2)}\notag\\
	\Rightarrow|\frac{e_{_{NL}}}{e_3}|=\frac{(\delta+2)\alpha}
	{(1+\delta)((\alpha+1)^2-2+\delta\alpha^2)}>1.\notag
	\end{align}
	\newline
	Case 2.4, if $e_1\times e_2<0$ and $e_3\times e_2>0$ then
	\begin{align}
	&e_{_{NL}}=\frac{\alpha^{-1}\times-(1+\delta)\alpha-1}{-\alpha^{-1}+(1+\delta)\alpha-2}e_2=
	\frac{-(\delta+2)\alpha}{(\alpha-1)^2-2+\delta\alpha^2}\notag\\
	&\Rightarrow|\frac{e_{_{NL}}}{e_3}|=\frac{(\delta+2)\alpha}
	{(1+\delta)((\alpha-1)^2-2+\delta\alpha^2)}>1\notag
	\end{align}
	where $0<\delta<\frac{1}{\alpha}-1$.
	
	In the case of super-linear convergent algorithms, it can be assumed that
	\begin{align}
	|\alpha_1|=\alpha\:,\:|\alpha_2|=(1-\delta)\alpha\quad;\quad 0<\delta<1
	\label{eq:nl_3}
	\end{align}
	which results in the same procedure as for the case of sub-linear convergence except for the sign of $\delta$.
	\newline
	Case 3.1, if all of the $e_i$'s have the same sign, then
	\begin{align}
	|\frac{e_{_{NL}}}{e_3}|=\frac{\delta}{(1-\delta)((\alpha-1)^2-\delta\alpha^2)}<1\:;\:\text{for}\: \delta<\delta_0
	\label{nl_3_1}
	\end{align}
	where $\delta_0=\frac{\alpha^2-\alpha+1-\sqrt{2\alpha^2-2\alpha+1}}{\alpha^2}$. Hence, the NL estimation diverges for $\delta_0<\delta<1$. It can be seen that for a cubically convergent sequence ($\delta=1-\alpha$), the NL estimation diverges since $\delta_0<1-\alpha$.
	\newline
	Case 3.2, if $e_1\times e_2<0$ and $e_3\times e_2<0$, then
	\begin{align}
	|\frac{e_{_{NL}}}{e_3}|=\frac{\delta}{(1-\delta)((\alpha+1)^2-\delta\alpha^2)}<1\,;\,\delta<\delta_0
	\label{nl_3_2}
	\end{align}
	where $\delta_0=\frac{\alpha^2+\alpha+1-\sqrt{2\alpha^2+2\alpha+1}}{\alpha^2}$. In this case, sequences with quadratic convergence can be accelerated if $\delta_0>1-\alpha$. In other words, these sequences can be accelerated if $0.36110<\alpha$.
	\newline
	Case 3.3, if $e_1\times e_2>0,e_3\times e_2<0$ and $0<\delta<1$ then
	\begin{align}
	|\frac{e_{_{NL}}}{e_3}|=\frac{(2-\delta)\alpha}
	{|(1-\delta)((\alpha+1)^2-2-\delta\alpha^2)|}>1.\notag
	\end{align}
	\newline
	Case 3.4, if $e_1\times e_2<0$ and $e_3\times e_2>0$ then
	\begin{align}
	|\frac{e_{_{NL}}}{e_3}|=\frac{(-\delta+2)\alpha}
	{|(1-\delta)((\alpha-1)^2-2-\delta\alpha^2)}|>1\:;\:0<\delta<1.\notag
	\end{align}
	
	\section*{References}
	
	\bibliography{mybibfile}
\end{document}